\documentclass[12pt,english]{article}
\usepackage{times}
\usepackage[T1]{fontenc}
\usepackage{geometry}
\usepackage{amsmath}
\usepackage{color}
\usepackage{graphicx}
\usepackage{setspace}
\usepackage{amssymb}

\makeatletter

\providecommand{\tabularnewline}{\\}

\usepackage{bm}

\usepackage{babel}
\makeatother
\begin{document}

\section*{\noindent Design of Clustered Phased Arrays by Means of an Innovative
Power Pattern Matching-Driven Method - The Linear Array Case}

\noindent \vfill

\noindent A. Benoni,$^{(1)(2)}$ \emph{Member, IEEE}, L. Poli,$^{(1)(2)}$
\emph{Member, IEEE}, P. Rocca, $^{(1)(2)(3)}$ \emph{Fellow, IEEE},
and A. Massa,$^{(1)(2)(4)(5)(6)}$ \emph{Fellow, IEEE}

\noindent \vfill

\noindent \textcolor{black}{\footnotesize ~}{\footnotesize \par}

\noindent \textcolor{black}{\scriptsize $^{(1)}$} \textcolor{black}{\emph{\scriptsize ELEDIA
Research Center}} \textcolor{black}{\scriptsize (}\textcolor{black}{\emph{\scriptsize ELEDIA}}\textcolor{black}{\scriptsize @}\textcolor{black}{\emph{\scriptsize UniTN}}
\textcolor{black}{\scriptsize - University of Trento)}{\scriptsize \par}

\noindent \textcolor{black}{\scriptsize DICAM - Department of Civil,
Environmental, and Mechanical Engineering}{\scriptsize \par}

\noindent \textcolor{black}{\scriptsize Via Mesiano 77, 38123 Trento
- Italy}{\scriptsize \par}

\noindent \textit{\textcolor{black}{\emph{\scriptsize E-mail:}}} {\scriptsize \{}\emph{\scriptsize arianna.benoni,
lorenzo.poli, paolo.rocca, andrea.massa}{\scriptsize \}}\textcolor{black}{\scriptsize @}\textcolor{black}{\emph{\scriptsize unitn.it}}{\scriptsize \par}

\noindent \textcolor{black}{\scriptsize Website:} \textcolor{black}{\emph{\scriptsize www.eledia.org/eledia-unitn}}{\scriptsize \par}

\noindent \textcolor{black}{\scriptsize ~}{\scriptsize \par}

\noindent \textcolor{black}{\scriptsize $^{(2)}$} \textcolor{black}{\emph{\scriptsize CNIT
- \char`\"{}University of Trento\char`\"{} ELEDIA Research Unit }}{\scriptsize \par}

\noindent \textcolor{black}{\scriptsize Via Sommarive 9, 38123 Trento
- Italy}{\scriptsize \par}

\noindent \textcolor{black}{\scriptsize Website:} \textcolor{black}{\emph{\scriptsize www.eledia.org/eledia-unitn}}{\scriptsize \par}

\noindent \textcolor{black}{\scriptsize ~}{\scriptsize \par}

\noindent {\scriptsize $^{(3)}$} \emph{\scriptsize ELEDIA Research
Center} {\scriptsize (}\emph{\scriptsize ELEDIA}{\scriptsize @}\emph{\scriptsize XIDIAN}
{\scriptsize - Xidian University)}{\scriptsize \par}

\noindent {\scriptsize P.O. Box 191, No.2 South Tabai Road, 710071
Xi'an, Shaanxi Province - China }{\scriptsize \par}

\noindent {\scriptsize E-mail:} \emph{\scriptsize paolo.rocca@xidian.edu.cn}{\scriptsize \par}

\noindent {\scriptsize Website:} \emph{\scriptsize www.eledia.org/eledia-xidian}{\scriptsize \par}

\noindent \textcolor{black}{\scriptsize ~}{\scriptsize \par}

\noindent \textcolor{black}{\scriptsize $^{(4)}$} \textcolor{black}{\emph{\scriptsize ELEDIA
Research Center}} \textcolor{black}{\scriptsize (}\textcolor{black}{\emph{\scriptsize ELEDIA}}\textcolor{black}{\scriptsize @}\textcolor{black}{\emph{\scriptsize UESTC}}
\textcolor{black}{\scriptsize - UESTC)}{\scriptsize \par}

\noindent \textcolor{black}{\scriptsize School of Electronic Science
and Engineering, Chengdu 611731 - China}{\scriptsize \par}

\noindent \textit{\textcolor{black}{\emph{\scriptsize E-mail:}}} \textcolor{black}{\emph{\scriptsize andrea.massa@uestc.edu.cn}}{\scriptsize \par}

\noindent \textcolor{black}{\scriptsize Website:} \textcolor{black}{\emph{\scriptsize www.eledia.org/eledia}}\textcolor{black}{\scriptsize -}\textcolor{black}{\emph{\scriptsize uestc}}{\scriptsize \par}

\noindent \textcolor{black}{\scriptsize ~}{\scriptsize \par}

\noindent \textcolor{black}{\scriptsize $^{(5)}$} \textcolor{black}{\emph{\scriptsize ELEDIA
Research Center}} \textcolor{black}{\scriptsize (}\textcolor{black}{\emph{\scriptsize ELEDIA@TSINGHUA}}
\textcolor{black}{\scriptsize - Tsinghua University)}{\scriptsize \par}

\noindent \textcolor{black}{\scriptsize 30 Shuangqing Rd, 100084 Haidian,
Beijing - China}{\scriptsize \par}

\noindent \textcolor{black}{\scriptsize E-mail:} \textcolor{black}{\emph{\scriptsize andrea.massa@tsinghua.edu.cn}}{\scriptsize \par}

\noindent \textcolor{black}{\scriptsize Website:} \textcolor{black}{\emph{\scriptsize www.eledia.org/eledia-tsinghua}}{\scriptsize \par}

\noindent \textcolor{black}{\footnotesize ~}{\footnotesize \par}

\noindent \textcolor{black}{\footnotesize $^{(6)}$} {\scriptsize School
of Electrical Engineering}{\scriptsize \par}

\noindent {\scriptsize Tel Aviv University, Tel Aviv 69978 - Israel}{\scriptsize \par}

\noindent \textit{\textcolor{black}{\emph{\scriptsize E-mail:}}} \emph{\scriptsize andrea.massa@eng.tau.ac.il}{\scriptsize \par}

\noindent \textcolor{black}{\scriptsize Website:} \textcolor{black}{\emph{\scriptsize https://engineering.tau.ac.il/}}{\scriptsize \par}

\noindent \textcolor{black}{\vfill}

\emph{This work has been submitted to the IEEE for possible publication.
Copyright may be transferred without notice, after which this version
may no longer be accessible.}

\newpage
\section*{Design of Clustered Phased Arrays by Means of an Innovative Power
Pattern Matching-Driven Method - The Linear Array Case}

~

\noindent ~

\noindent ~

\begin{flushleft}A. Benoni, L. Poli, P. Rocca, and A. Massa\end{flushleft}

\textcolor{red}{\vfill}

\begin{abstract}
The design of sub-arrayed phased arrays (\emph{PA}s) with sub-array-only
amplitude and phase controls that afford arbitrary-shaped power patterns
matching reference ones is addressed. Such a synthesis problem is
formulated in the power pattern domain and an innovative complex-excitations
clustering method, which is based on the decomposition of the reference
power pattern in a number of elementary patterns equal to the array
elements, is presented. A set of representative results is reported
to illustrate the features of the proposed approach as well as to
assess its effectiveness in comparison with benchmark results from
the state-of-the-art (\emph{SoA}) excitation matching-based clustering
methods.

\noindent \textcolor{red}{\vfill}
\end{abstract}
\noindent \textbf{Key words}: Phased Array, Linear Array, Clustered
Array, Power Pattern Matching, k-means Algorithm, Iterative Projection
Method.

\newpage
\section{Introduction}

Nowadays, phased array (\emph{PA}) antennas are widely employed in
different fields such as radar, mobile communications, biomedical,
space, and optical systems \cite{Haupt 2015}-\cite{Woo 2022}. Thanks
to a fully electronic control of the radiation pattern, modern \emph{PA}s
allow an agile reconfiguration as well as wide beam scanning capabilities,
which are frequently mandatory in several applicative scenarios \cite{Mailloux 2018}.
On the other hand, the use of a transmit/receive module (\emph{TRM})
at each array element to set the amplitude and the phase/time-delay
of either the transmitted or the received signal for beamforming purposes,
implies high costs \cite{Herd 2016}. This issue is an impairment
for the large-scale deployment in mass-market commercial applications.
To overcome this drawback, different unconventional \emph{PA} architectures
have been proposed \cite{Rocca 2016} such as sub-arrayed/clustered,
thinned \cite{Skolnik 1964}\cite{Keizer 2008}, or sparse \cite{Goudos 2011}\cite{Liu 2008}
layouts. In clustered phased arrays (\emph{CPA}s), a single \emph{TRM}
is shared among multiple antenna elements. However, simply partitioning
the antenna aperture into regular sub-arrays of equal size and orientation
is not effective because of the occurrence of undesired grating lobes
as the scanning and the bandwidth requirements increase \cite{Haupt 2010}-\cite{Balanis 2016}.
Therefore, irregular/a-periodic sub-arrayed arrangements have been
proposed to reduce quantization effects and grating lobes also jointly
guaranteeing a high aperture efficiency unlike thinned or sparse array
layouts \cite{Haupt 1985}\cite{Rocca 2015}.

\noindent Designing \emph{CPA}s implies the solution of two sub-problems,
namely the clustering problem and the weighting one. The clustering
problem is aimed at grouping the array elements into a set of mutually
exclusive and exhaustive sub-sets. The weighting problem goal is that
of defining, for a given clustering, the complex (i.e., amplitude
and phase) sub-array excitations to fulfil user-defined specifications.
The grouping of the array elements into clusters is a hard combinatorial
problem whatever the objective (e.g., matching a reference solution
or optimizing selected pattern features). For instance, the cardinality
of the solution space of possible clustering configurations for a
linear array grows exponentially with the number of array elements
according to the Stirling number of the second kind \cite{Abramowitz 1972}.
Since an enumerative solution strategy would be prohibitive even for
small/medium arrays, different methodologies based on deterministic
\cite{McNamara 1988}\cite{Xiong 2013}, stochastic \cite{Goudos 2013}-\cite{Yang 2021},
and hybrid \cite{D'Urso 2007}\cite{Lopez 2001} techniques have been
proposed in the state-of-the-art (\emph{SoA}) literature to determine
the best sub-array architecture and set of cluster excitations fitting
the user-defined constraints/requirements. Although very efficient,
these methods have been successfully applied only to small- and medium-size
arrays owing to the high computational cost. By casting the \emph{CPA}
design within the excitation matching (\emph{EM}) framework (i.e.,
defining the sub-array weights to match a set of reference excitations,
one for each array element) and exploiting the Fisher's grouping theory
\cite{Fisher 1958}, the cardinality of the solution space has been
reduced to the set of contiguous partitions of the ordered list of
reference excitations. The size of this latter space grows as the
binomial of the number of elements and it has been sampled with the
\emph{Contiguous Partition Method} (\emph{CPM}) \cite{Manica 2008}-\cite{Rocca 2019}.
Different customizations of the \emph{CPM} have been derived to deal
with different array architectures considering either amplitude-only
\cite{Manica 2008}-\cite{Rocca 2009} or phase-only \cite{Rocca 2019}
sub-array control. Fully sub-arrayed architectures with a joint control
of the amplitude and the phase have not been dealt with the \emph{CPM}
since the Fisher's grouping theory guarantees the best partitioning,
in the least-square sense, of a set of reference excitations only
when real-valued. To overcome such a \emph{CPM} limitation, the clustering
of \emph{PA}s with complex sub-array excitations has been reformulated
as an optimization one then solved \cite{Rocca 2020}\cite{Rocca 2022}
with a customized version of the k-means \cite{Arthur 2009}\cite{Battiti 2017}.
More in detail, the array clustering has been defined in the two-dimensional
Gauss plane of the complex excitations, while the sub-array weights
have been computed in closed-form as the arithmetic mean of the reference
excitations belonging to a cluster \cite{Manica 2008}\cite{Rocca 2019}.
The \emph{EM} k-means has been chosen because of the computational
efficiency of the k-means algorithm \cite{Arthur 2009} and for being
the natural extension of the Fisher's theory \cite{Fisher 1958} to
two-dimensional/complex domains. However, it should be pointed out
that the \emph{EM} k-means method yields the best sub-array configuration
by solving an \emph{EM} problem without guaranteeing the optimal matching
with the power pattern generated by the reference excitations. A first
attempt to overcome this issue has been presented in \cite{Benoni 2022}
where the solution space of the (complex) reference excitations has
been sampled to find the \emph{PA} clustering that minimizes the power
pattern matching (\emph{PM}) metric. More specifically, the dimension
of the clustering problem has been first reduced from the two-dimensional
complex-space down to a one-dimensional real-space by exploiting the
theory of the space filling curves \cite{Sagan 1994}. Then, a customized
clustering algorithm, namely the \emph{Swap Element Algorithm} (\emph{SEA}),
has been proposed to determine the optimal grouping of the array elements
\cite{Benoni 2022}. 

\noindent Within the \emph{PM} framework, this paper proposes, for
the first time to the best of the authors' knowledge, an approach
for addressing the clustering problem directly in the power pattern
domain. Towards this end, the power pattern of a reference fully populated
\emph{PA} (\emph{FPA}) is first decomposed into {}``\emph{Elementary
Power Patterns}'' (\emph{EP}s), one for each array element. The k-means
algorithm is then applied in the power pattern domain to yield the
best sub-array configuration by minimizing the \emph{PM} metric. Once
the clustering is defined, the sub-array weights are finally computed
with a customized version of the \emph{Iterative Projection Method}
(\emph{IPM}) \cite{Bucci 1990}.

\noindent The main novelties of this work over the existing \emph{SoA}
literature comprise (\emph{i}) the mathematical formulation of an
innovative paradigm for the synthesis of \emph{CPA}s whose power patterns
maximize the matching with reference ones, (\emph{ii}) the implementation
of a customized clustering method working in the power pattern space
instead of in the excitation one as generally done in \emph{SoA} \emph{PA}
clustering methods; (\emph{iii}) the design of \emph{CPA}s with sub-array-only
amplitude and phase excitations that outperform \emph{SoA} methods
in matching reference power patterns.

\noindent The remaining of this paper is organized as follows. The
mathematical formulation of the \emph{PA} clustering directly in the
\emph{PM} framework, including the definition of the \emph{EP}s, is
described in Sect. 2, while the proposed \emph{PM}-driven \emph{CPA}
synthesis method is presented in Sect. 3. Section 4 reports a set
of representative numerical results to assess the effectiveness of
the proposed clustering method also in a comparative fashion. Finally,
some conclusions are drawn (Sect. 5).

\section{\noindent Mathematical Formulation}

Let us consider a linear \emph{PA} of $N$ isotropic elements disposed
along the \emph{x-}axis with inter-element spacing $d=\frac{\lambda}{2}$.
The array elements are grouped into $Q$ ($Q<N$) clusters, each containing
$N_{q}$ ($q=1,...,Q$) elements so that $\sum_{q=1}^{Q}N_{q}=N$.
Every $q$-th ($q=1,...,Q$) cluster has a single \emph{TRM} composed
by an amplifier and a phase shifter providing an amplitude and a phase
equal to $\alpha_{q}$ and $\varphi_{q}$, respectively (Fig. 1).
The power pattern ($P\left(u\right)\triangleq\left|AF\left(u\right)\right|^{2}$;
$AF\left(u\right)$ being the array factor given by $AF\left(u\right)=\sum_{q=1}^{Q}I_{q}\sum_{n=1}^{N}\delta_{c_{n}q}e^{jk\left(n-1\right)du}$)
of such a \emph{CPA} is\begin{equation}
P\left(u\right)=\left|\sum_{q=1}^{Q}I_{q}\sum_{n=1}^{N}\delta_{c_{n}q}e^{jk\left(n-1\right)du}\right|^{2}\label{eq:_CA.Pattern}\end{equation}
where $I_{q}=\alpha_{q}e^{j\varphi_{q}}$ ($q=1,...,Q$) and $\delta_{c_{n}q}$
is the Kronecker delta function equal to $\delta_{c_{n}q}=1$ if the
$n$-th ($n=1,...,N$) element belongs to the $q-$th cluster (i.e.,
$c_{n}=q$) and to $\delta_{c_{n}q}=0$, otherwise. Moreover, $k=\frac{2\pi}{\lambda}$
is the wave-number, $\lambda$ being the wavelength at the \emph{CPA}
working frequency, and $u=\sin\theta$, $\theta$ being the angular
variable computed from the direction orthogonal to the array axis
($\theta=\left[-90,\ 90\right]$ {[}deg{]}).

\noindent To synthesize such a \emph{CPA}, the following {}``\emph{PM-Driven
CPA Design Problem}\textbf{\emph{''}} is solved

\begin{quotation}
\noindent Given a set of $N$ reference complex excitations \{$I_{n}$;
$n=1,...,N$\} ($I_{n}=\alpha_{n}e^{j\varphi_{n}}$) of a \emph{FPA}
affording the reference power pattern {[}$P^{ref}\left(u\right)\triangleq\left|AF^{ref}\left(u\right)\right|^{2}${]}\begin{equation}
P^{ref}\left(u\right)\triangleq\left|\sum_{n=1}^{N}I_{n}e^{jkd\left(n-1\right)u}\right|^{2},\label{eq:_FPA.Pattern}\end{equation}
 determine the clustering vector, \emph{}$\mathbf{c}^{opt}=\left\{ c_{n}^{opt}\in\left[1:Q\right];\, n=1,...,N\right\} $,
which univocally describes the grouping of the $N$ array elements
into $Q$ clusters, and the corresponding sub-array weights, $\mathbf{I}^{opt}=\left\{ I_{q}^{opt};\, q=1,...,Q\right\} $,
that minimize the \emph{PM} metric defined as\begin{equation}
\Gamma\left(\mathbf{c},\ \mathbf{I}\right)=\frac{\int_{-1}^{1}\left|P^{ref}\left(u\right)-P\left(u;\ \mathbf{c},\ \mathbf{I}\right)\right|du}{\int_{-1}^{1}P^{ref}\left(u\right)du},\label{eq:_PM.Metric}\end{equation}
which quantifies the mismatch between the power patterns radiated
by the reference \emph{FPA} (\ref{eq:_FPA.Pattern}) and the \emph{CPA}
(\ref{eq:_CA.Pattern}).
\end{quotation}

\section{\emph{PM}-Driven \emph{CPA} Design Method\label{sub:PM-K-means-Method}}

Unlike \emph{CPA} synthesis methods developed in the \emph{EM} framework
\cite{Manica 2008}-\cite{Rocca 2022}, where the grouping of the
$N$ array elements is done in the space of the (complex) excitations,
$\mathcal{S}_{I}$, by minimizing the mismatch between the reference
excitations and the sub-arrayed ones, the \emph{CPA} design problem
is addressed here in the power pattern domain, $\mathcal{S}_{P}$
(Fig. 2) according to the iterative procedure presented in the following
after some premises. 

\noindent Let us consider the $n$-th ($n=1,...,N$) term of the array
factor of the reference \emph{FPA}, $AF^{ref}\left(u\right)$,

\begin{equation}
AF_{n}\left(u\right)=I_{n}e^{jkd\left(n-1\right)u},\label{eq:_ElementaryAF}\end{equation}
its power pattern (\ref{eq:_FPA.Pattern}) can be rewritten as\\
\begin{equation}
P^{ref}\left(u\right)=\sum_{n=1}^{N}\left[\left|AF_{n}\left(u\right)\right|^{2}\sum_{\ell=1,\ \ell\neq n}^{N}AF_{n}\left(u\right)AF_{\ell}^{*}\left(u\right)\right]\label{eq:_FPA.Pattern.Decomposed}\end{equation}

where the superscript $^{*}$ stands for complex conjugate. According
to (\ref{eq:_FPA.Pattern.Decomposed}), the reference power pattern
turns out to be the linear combination of $N$ \emph{EP}s, the $n$-th
($n=1,...,N$) one given by\begin{equation}
P_{n}\left(u\right)=\left|AF_{n}\left(u\right)\right|^{2}+\sum_{\ell=1,\ \ell\neq n}^{N}AF_{n}\left(u\right)AF_{\ell}^{*}\left(u\right),\label{eq:_Elementary.Power.Pattern}\end{equation}
so that\begin{equation}
P^{ref}\left(u\right)=\sum_{n=1}^{N}P_{n}\left(u\right).\label{eq:_FPA.Pattern.w.Elementary.Power.Patterns}\end{equation}
It is worth noting that, by definition (\ref{eq:_Elementary.Power.Pattern}),
the $n$-th ($n=1,...,N$) \emph{EP,} $P_{n}\left(u\right)$, does
not only depend on the corresponding $n$-th array element, but it
is composed by an additional term that takes into account the cross-correlation
of the $n$-th element with all the other $N-1$ array elements. Moreover,
$P_{n}\left(u\right)$ ($n=1,...,N$) is a complex quantity, but the
summation of the $N$ \emph{EP}s (\ref{eq:_FPA.Pattern.w.Elementary.Power.Patterns})
is a real-valued quantity (see \emph{Appendix}) equal to the array
power pattern of the \emph{FPA}.

\noindent Starting from these considerations, the rationale behind
the proposed clustering strategy is to group, within the same sub-array,
the elements having similar \emph{EP}, $P_{n}\left(u\right)$ ($n=1,...,N$).
However, unlike \emph{EM}-driven clustering techniques, where the
reference excitations are scalar variables, the \emph{PM}-driven \emph{CPA}
synthesis method (\emph{PMM}) deals with \emph{EP}s that are continuous
functions of the angular variable $u$ ($u\in\left[-1,\ 1\right]$).
Accordingly, the \emph{CPA} design problem is addressed within the
\emph{PM} framework by following a two-step iterative procedure where
the first step ({}``\emph{Clustering Step}'') is aimed at defining
a trial array clustering by means of a customized \emph{PM}-driven
k-means method, while the sub-array excitations are optimized in the
second step ({}``\emph{Weighting Step}'') with the \emph{IPM} until
the convergence of the synthesis process.

\noindent More specifically, the two-step procedure is implemented
as shown in Fig. 3 and detailed hereinafter:

\begin{itemize}
\item \textbf{Step 0} - \emph{Reference Pattern Selection}. Given the array
geometry (i.e., $N$ and $d$), choose the desired reference power
pattern, $P^{ref}\left(u\right)$, and input the corresponding set
of $N$ reference complex excitations, \{$I_{n}$; $n=1,...,N$\};
\item \textbf{Step 1 (}\textbf{\emph{Clustering Step}}\textbf{)} - \emph{EPs
Definition}. Compute the \emph{EP}s, \{$P_{n}\left(u\right)$; $n=1,...,N\}$,
according to (\ref{eq:_Elementary.Power.Pattern}) and uniformly discretize
the angular domain into $M$ samples, \{$u_{m}$; $m=1,...,M\}$,
being $u_{m}=-1+\frac{2\left(m-1\right)}{M-1}$. Set $m=1$;

\begin{itemize}
\item \textbf{Step 1.1 (}\textbf{\emph{Clustering Step}}\textbf{)} - \emph{Centroids
Initialization}. Set $r=0$, $r$ being the iteration index of the
k-means procedure and normalize the values of the \emph{EP}s sampled
at the $m$-th angular direction \begin{equation}
\tilde{P}_{n}\left(u_{m}\right)=\frac{P_{n}\left(u_{m}\right)}{{\displaystyle \max_{n=1,...,N}}\left\{ \left|P_{n}\left(u_{m}\right)\right|\right\} }.\label{eq:_normalized.EPs}\end{equation}
Randomly select the initial $Q$ centroids, \{$W_{q}^{\left(r\right)}\left(u_{m}\right)$;
$q=1,...,Q\}$, among the $N$ available normalized \emph{EP} values,
\{$\tilde{P}_{n}\left(u_{m}\right)$; $n=1,...,N\}$;
\item \textbf{Step 1.2 (}\textbf{\emph{Clustering Step}}\textbf{)} - \emph{Distance
Computation}. For each $n$-th ($n=1,...,N$) \emph{EP}, $\tilde{P}_{n}\left(u_{m}\right)$,
compute the Euclidean distance from the $q$-th ($q=1,...,Q$) centroid,
$W_{q}^{\left(r\right)}\left(u_{m}\right)$, \begin{eqnarray}
\xi_{nq}^{\left(r\right)}\left(u_{m}\right) & = & \left\Vert W_{q}^{\left(r\right)}\left(u_{m}\right)-\tilde{P}_{n}\left(u_{m}\right)\right\Vert \nonumber \\
 & = & \left[\left(\Re\left\{ W_{q}^{\left(r\right)}\left(u_{m}\right)\right\} -\Re\left\{ \tilde{P}_{n}\left(u_{m}\right)\right\} \right)^{2}+\right.\label{eq:_Euclidean.Distance}\\
 &  & +\left.\left(\Im\left\{ W_{q}^{\left(r\right)}\left(u_{m}\right)\right\} -\Im\left\{ \tilde{P}_{n}\left(u_{m}\right)\right\} \right)^{2}\right]^{\frac{1}{2}}\nonumber \end{eqnarray}
$\Re\left\{ \cdot\right\} $ and $\Im\left\{ \cdot\right\} $ being
the real part and the imaginary one, respectively;
\item \textbf{Step 1.3 (}\textbf{\emph{Clustering Step}}\textbf{)} - \emph{Element
Clustering}. Determine the $r$-th clustering vector, $\mathbf{c}_{m}^{\left(r\right)}=\mathbf{c}^{\left(r\right)}\left(u_{m}\right)$,
by associating each $n$-th ($n=1,...,N$) normalized \emph{EP}, $\tilde{P}_{n}\left(u_{m}\right)$,
to the $q$-th ($q=1,...,Q$) cluster (i.e., $c_{n}^{\left(k\right)}\left(u_{m}\right)=q$)
being
\end{itemize}
\begin{equation}
q=arg\left\{ \min_{p\in\left[1,\, Q\right]}\left[\xi_{np}^{\left(r\right)}\left(u_{m}\right)\right]\right\} \,;\label{eq:_Cluster.Assignment}\end{equation}

\begin{itemize}
\item \textbf{Step 1.4 (}\textbf{\emph{Clustering Step}}\textbf{)} - \emph{Centroids
Update}. Update the k-means iteration index, $r\leftarrow r+1$, and
the value of the $q$-th ($q=1,...,Q$) centroid as follows\begin{equation}
W_{q}^{\left(r\right)}=\frac{\sum_{n=1}^{N}\delta_{c_{n}^{\left(r-1\right)}\left(u_{m}\right)q}\tilde{P}_{n}\left(u_{m}\right)}{N_{q}^{\left(r-1\right)}\left(u_{m}\right)}\label{eq:_Cetroids}\end{equation}
where $N_{q}^{\left(r-1\right)}\left(u_{m}\right)=\sum_{n=1}^{N}\delta_{c_{n}^{\left(r-1\right)}\left(u_{m}\right)q}$;
\item \textbf{Step 1.5 (}\textbf{\emph{Clustering Step}}\textbf{)} - \emph{k-means
Convergence Check.} If the index $r$ is greater than the maximum
number of iterations $R$ (i.e., $r>R$) or the stationary condition,
$W_{q}^{\left(r\right)}=W_{q}^{\left(r-1\right)},\ \forall q\in\left[1,\  Q\right]$,
is reached, then stop the iterative k-means procedure, set the optimal
clustering configuration to $\mathbf{c}_{m}^{opt}=\mathbf{c}_{m}^{\left(r-1\right)}$,
$N_{q}\left(u_{m}\right)=N_{q}^{\left(r-1\right)}\left(u_{m}\right)$
($q=1,...,Q$), and go to Step 2. Otherwise, repeat Step 1.2;
\end{itemize}
\item \textbf{Step 2 (}\textbf{\emph{Weighting Step}}\textbf{)} - \emph{Excitations
Initialization.} Set $t=0$, $t$ being the iteration index of the
\emph{IPM}, and initialize the set of $N$ auxiliary excitations \{$I_{n}^{\left(t\right)}\left(u_{m}\right)$;
$n=1,...,N\}$, to the corresponding ones of the reference \emph{FPA}:
$\left.I_{n}^{\left(t\right)}\left(u_{m}\right)\right\rfloor _{t=0}=I_{n}$
($n=1,...,N$); 

\begin{itemize}
\item \textbf{Step 2.1 (}\textbf{\emph{Weighting Step}}\textbf{)} - \emph{Sub-array
Weights Computation.} Compute the $q$-th sub-array excitations as\begin{equation}
I_{q}^{\left(t\right)}\left(u_{m}\right)=\frac{\sum_{n=1}^{N}\delta_{c_{n}^{opt}\left(u_{m}\right)q}I_{n}^{\left(t\right)}\left(u_{m}\right)}{N_{q}\left(u_{m}\right)};\label{eq:_Subarray.Excitations.IPM}\end{equation}

\item \textbf{Step 2.2 (}\textbf{\emph{Weighting Step}}\textbf{)} - \emph{Projection
onto Reference Power Pattern.} Compute through (\ref{eq:_CA.Pattern})
the power pattern, $P_{m}^{\left(t\right)}\left(u\right)=P\left(u;\ \mathbf{c}_{m}^{opt},\ \mathbf{I}_{m}^{\left(t\right)}\right)$,
of the \emph{CPA} with excitation vector $\ \mathbf{I}_{m}^{\left(t\right)}=\left\{ I_{q}^{\left(t\right)}\left(u_{m}\right);\negmedspace q=1,...,Q\right\} $.
Determine the projected power pattern, $\hat{P}_{m}^{\left(t\right)}\left(u\right)$
($\hat{P}_{m}^{\left(t\right)}\left(u\right)\triangleq\left|\hat{AF}_{m}^{\left(t\right)}\left(u\right)\right|^{2}$)
by projecting $P_{m}^{\left(t\right)}\left(u\right)$ onto $P^{ref}\left(u\right)$.
More in detail, for each $l$-th ($l=1,...,M$) angular sample, set\begin{equation}
\hat{AF}_{m}^{\left(t\right)}\left(u_{l}\right)=\frac{\sum_{q=1}^{Q}I_{q}^{\left(t\right)}\left(u_{m}\right)\sum_{n=1}^{N}\delta_{c_{n},q}e^{jk\left(n-1\right)du_{l}}}{\left|\sum_{q=1}^{Q}I_{q}^{\left(t\right)}\left(u_{m}\right)\sum_{n=1}^{N}\delta_{c_{n},q}e^{jk\left(n-1\right)du_{l}}\right|}\times\sqrt{P^{ref}\left(u_{l}\right)}\label{eq:_Projection.on.Mask}\end{equation}
 if $P_{m}^{\left(t\right)}\left(u_{l}\right)>P^{ref}\left(u_{l}\right)$
or $P_{m}^{\left(t\right)}\left(u_{l}\right)<P^{ref}\left(u_{l}\right)$,
while $\hat{AF}_{m}^{\left(t\right)}\left(u_{l}\right)$ $=$ $\sum_{q=1}^{Q}$
$I_{q}^{\left(t\right)}\left(u_{m}\right)$ $\sum_{n=1}^{N}$ $\delta_{c_{n}^{opt}\left(u_{m}\right)q}e^{jk\left(n-1\right)du_{l}}$,
otherwise;
\item \textbf{Step 2.3 (}\textbf{\emph{Weighting Step}}\textbf{)} - \emph{Fitness
Evaluation}. Evaluate the \emph{PM} metric (\ref{eq:_PM.Metric})
for the projected power pattern, $\hat{P}_{m}^{\left(t\right)}\left(u\right)$,
\begin{equation}
\Gamma_{m}^{\left(t\right)}=\Gamma\left(\ \mathbf{c}_{m}^{opt},\ \mathbf{I}_{m}^{\left(t\right)}\right)=\frac{\int_{-1}^{1}\left|P^{ref}\left(u\right)-\hat{P}_{m}^{\left(t\right)}\left(u\right)\right|du}{\int_{-1}^{1}P^{ref}\left(u\right)du};\label{eq:_PM.Metric.IPM}\end{equation}

\item \textbf{Step 2.4 (}\textbf{\emph{Weighting Step}}\textbf{)} - \emph{IPM
Convergence Check.} Stop the iterative \emph{IPM} procedure if the
index $t$ is greater than the maximum number of iterations $T$ (i.e.,
$t>T$) or the value of the \emph{PM} metric (\ref{eq:_PM.Metric.IPM})
is stationary (i.e., $\Gamma_{m}^{\left(t\right)}=\Gamma_{m}^{\left(t-1\right)}$),
then set the optimal excitations to $\mathbf{I}_{m}^{opt}=\mathbf{I}_{m}^{\left(t\right)}$
and jump to Step 3. Otherwise, go to Step 2.5;
\item \textbf{Step 2.5} \textbf{(}\textbf{\emph{Weighting Step}}\textbf{)}
- \emph{Projection onto Excitations Space}. Derive a new set of auxiliary
excitations, \{$I_{n}^{\left(t+1\right)}\left(u_{m}\right)$; $n=1,...,N\}$,
through the inverse Fourier transform of $\hat{AF}_{m}^{\left(t\right)}\left(u\right)$,
update the iteration index, $t\leftarrow t+1$, and repeat Step 2.1;
\end{itemize}
\item \textbf{Step 3} - \textbf{}\emph{Convergence Check}. If $m\neq M$
then update the angular direction index, $m\leftarrow m+1$, and return
to Step 1.1;
\item \textbf{Step 4} - \textbf{}\emph{Optimal Solution} \emph{Output}.
Set the optimal clustering and excitations to the trial solution with
the best \emph{PM} metric, $\Gamma^{opt}$,\begin{equation}
\left(\mathbf{c}^{opt},\ \mathbf{I}^{opt}\right)=arg\left\{ \min_{m\in\left[1,\, M\right]}\Gamma\left(\ \mathbf{c}_{m}^{opt},\ \mathbf{I}_{m}^{opt}\right)\right\} .\label{eq:_Optimal.Solution}\end{equation}

\end{itemize}

\section{Numerical Results}

In this section, a set of representative results is presented to give
the interested readers some insights on the effectiveness of the proposed
\emph{PMM} in comparison with \emph{SoA} clustering techniques, as
well.

\noindent The first example is aimed at describing the behavior of
the \emph{PMM} and it deals with a sub-arrayed linear array of $N=12$
elements spaced by $d=\frac{\lambda}{2}$ with $Q=8$ (i.e., $Q=\frac{3N}{4}$)
clusters. The \emph{CPA} layout and the corresponding sub-array weights
have been optimized with the \emph{PMM} to radiate a pattern as close
as possible to that of a reference Dolph-Chebyshev (\emph{DC}) \cite{Mailloux 2018}
\emph{FPA} characterized by a pencil beam steered along the angular
direction $\theta_{0}^{ref}=10$ {[}deg{]} and with $SLL^{ref}=-20$
{[}dB{]}. For illustrative purposes, $M=17$ angular samples have
been taken into account (Fig. 3). Moreover, since the performance
of the k-means algorithm depends on the initialization \cite{Arthur 2009},
it has been run $\sigma=50$ times, each with a different random seed,
to address the \emph{{}``Clustering Step}''.

\noindent Figure 4 shows the distributions of the $N=12$ \emph{EP}s,
\{$\tilde{P}_{n}\left(u_{m}\right)$; $n=1,...,N\}$, (Step 1.1) at
the angular samples $u_{m}=\left\{ 0.00,\ 0.25,\ 0.50,\ 0.75\right\} $,
while the intermediate results of the \emph{PMM} are summarized in
Fig. 5. More specifically, the best $u_{m}$-th clusterings yielded
after the $\sigma$ k-means runs are given in Figs. 5(\emph{a})-(\emph{d}),
while the corresponding sub-array groupings along the array layouts
are reported in Figs. 5(\emph{e})-(\emph{h}). For each angular sample,
$u_{m}$ ($m=1,...,M$), the clustering configuration, $\mathbf{c}_{m}^{opt}$
(Step 1.5), and thus the \emph{IPM}-computed sub-array excitations,
$\mathbf{I}_{m}^{opt}$ {[}Figs. 5(\emph{i})-(\emph{n}){]} (Step 2.5),
generally turn out different. For completeness, the corresponding
power patterns are shown in Fig. 6(\emph{a}), while Figure 6(\emph{b})
reports the value of the \emph{PM} metric for all the $M=17$ angular
samples. As it can be inferred {[}Fig. 6(\emph{b}){]}, the optimal
solution is that at $u=0.00$ where $\Gamma$ is minimum ($\Gamma^{opt}=5.94\times10^{-2}$).

\noindent To assess the effectiveness of the \emph{PMM} in finding
the optimal \emph{CPA} architecture, all possible $T=159027$ (i.e.,
the Stirling number for $N=12$ and $Q=8$) array-elements aggregations,
have been evaluated by determining, for each clustering, the sub-array
excitations through the \emph{IPM} and then computing the corresponding
\emph{PM} metric. Among the $T$ clustering configurations, no one
is better (in terms of \emph{PM} metric) than the one obtained by
the \emph{PMM} and only twice the same optimal/minimum pattern matching
value ($\Gamma^{opt}=5.94\times10^{-2}$) has been found, the patterns
yielded by the enumerative \emph{PM} analysis (\emph{EPM}) and synthesized
with the \emph{PMM} being identical (Fig. 7).

\noindent To further check the reliability of the proposed approach
to converge to the optimal solution, the second set of test cases
considers first the same reference pattern of the previous one, but
reducing the number of sub-arrays to $Q=\frac{N}{2}$. Moreover, dealing
with both the same array geometry (i.e., $N=12$ and $d=\frac{\lambda}{2}$)
and number of sub-arrays (i.e., $Q=\left\{ \frac{1}{2},\,\frac{3}{4}\right\} \times N$),
the \emph{PMM} has been applied to synthesize a \emph{CPA} matching
a Taylor pencil beam with $\theta_{0}^{ref}=10$ {[}deg{]}, $SLL^{ref}=-20$
{[}dB{]}, and $\overline{n}=3$. In all cases, $M=1001$ angular samples
have been now considered.

\noindent After running $\sigma=50$ times the k-means (\emph{PMM}
- Step 1), the global optimal solution has been reached in all cases
with an occurrence higher than 10\% (Fig. 8). Moreover, the analysis
of the behavior of the \emph{PM} metric as function of $u_{m}$ ($m=1,...,M$)
(Fig. 9) shows that the optimal clustered layout is always found in
correspondence with angular samples belonging to the main-lobe of
the reference pattern, which is identified in Fig. 9 by the first-null
beam-width (\emph{FNBW}) region.

\noindent The objective of the third numerical assessment is twofold.
Firstly, it is devoted to test the \emph{PMM} when dealing with larger
arrays with a number of elements, $N$, making impractical/unfeasible
the use of the \emph{EPM}. The second aim is to compare the performance
of the \emph{PMM} with those of a competitive \emph{SoA} \emph{EM}
method (\emph{EMM}) \cite{Rocca 2020}. Towards this end, let us consider
a set of array layouts with $N=\left\{ 16,\ 32,\ 48,\ 64\right\} $
and $Q=\left\{ \frac{1}{2},\,\frac{3}{4}\right\} \times N$, all targeting
a reference \emph{DC} pattern with $\theta_{0}^{ref}=10$ {[}deg{]}
and $SLL^{ref}=-20$ {[}dB{]}. To compare the \emph{EMM} \cite{Rocca 2020}
and the \emph{PMM} solutions, the matching improvement index, $\mathcal{R}$,
defined as\begin{equation}
\mathcal{R}=\frac{\Gamma^{EMM\ }-\Gamma^{PMM}}{\Gamma^{EMM}}\times100\label{eq:_Matching.Index}\end{equation}
has been used. In (\ref{eq:_Matching.Index}), $\Gamma^{EMM}$ and
$\Gamma^{PMM}$ are the values of the \emph{PM} metric for the optimal
solutions of the \emph{EMM} and the \emph{PMM}, respectively. It turns
out that the \emph{PMM} overcomes the \emph{EMM} when $\mathcal{R}>0$
and vice versa when $\mathcal{R}<0$.

\noindent The advantage of addressing the clustering problem in the
power pattern domain can be easily inferred from Fig. 10. Indeed,
the \emph{PMM} always outperforms the \emph{EMM} with a matching improvement
larger than $\mathcal{R}\geq30\%$. For illustrative purposes, the
optimal power patterns synthesized with the \emph{PMM} and the \emph{EMM}
are compared with the reference solution in Fig. 11 for the representative
cases of $N=32$ elements and either $Q=16$ ($Q=\frac{N}{2}$) {[}Fig.
11(\emph{a}){]} or $Q=24$ ($Q=\frac{3N}{4}$) {[}Fig. 11(\emph{b}){]}
sub-arrays. Although the relative improvement (\ref{eq:_Matching.Index})
is greater when $Q=\frac{3N}{4}$ (i.e., $\left.\mathcal{R}\right\rfloor _{Q=\frac{3N}{4}}=0.62>\left.\mathcal{R}\right\rfloor _{Q=\frac{N}{2}}=0.49$),
it is also worth noticing that the bigger improvement, in terms of
absolute \emph{PM} index (\ref{eq:_PM.Metric}), arises when $\frac{Q}{N}$
gets smaller (i.e., $\left.\frac{\Gamma^{EMM\ }}{\Gamma^{PMM}}\right\rfloor _{Q=\frac{3N}{4}}=2.64<\left.\frac{\Gamma^{EMM\ }}{\Gamma^{PMM}}\right\rfloor _{Q=\frac{N}{2}}=1.97$)
making the use of the \emph{PMM} very attractive.

\noindent The last example is concerned with the generation of shaped
beams instead of pencil beams as in the previous benchmark cases.
More specifically, a cosecant-squared (\emph{CS}) power pattern with
$SLL^{ref}=-20$ {[}dB{]}, main-lobe ripple $RPE=1.0$ {[}dB{]}, and
$FNBW=40$ {[}deg{]} has been considered as reference and it has been
assumed to be radiated by a linear array with $N=32$ elements. For
the \emph{CPA}, the number of sub-arrays has been set to $Q=16$ ($Q=\frac{N}{2}$).

\noindent Figure 12 compares the power patterns {[}Fig. 12(\emph{a}){]}
and the clustered array configurations {[}Figs. 12(\emph{b})-(\emph{e}){]}
synthesized with the \emph{PMM} and the \emph{EMM.} Also in this case,
the \emph{PMM} pattern better matches (Tab. I) the reference one by
improving the \emph{EMM} performance of $\mathcal{R}=51\%$ and yielding
a side-lobe level closer to the reference value for an amount of $2.25$
{[}dB{]} (Tab. I). To extensively confirm the superiority of the \emph{PMM}
over the \emph{EMM}, the \emph{CPA} process has been carried out by
varying the side-lobe level (i.e., $SLL^{ref}=\left\{ -20,\ -25,\ -30,\ -35,\ -40\right\} $
{[}dB{]}) (Figs. 13-14) and the steering angle (i.e., $\theta_{0}^{ref}=\left\{ 0,\ 5,\ 10,\ 15,\ 20\right\} $)
(Figs. 15-16), while keeping unaltered the shape of the main-lobe,
but changing the clustering ratios (i.e., $Q=\left\{ \frac{1}{4},\,\frac{1}{2},\,\frac{3}{4}\right\} \times N$).

\noindent From the analysis of Fig. 13 and Fig. 15, it turns out that
$\mathcal{R}>0$ regardless the value of \emph{SLL}, $\theta_{0}$,
or $Q$ of the scenario at hand. For the sake of comparison, both
\emph{PMM} and \emph{EMM} power patterns when $SLL=-25$ {[}dB{]}
(Fig. 14 - Tab. II) and $\theta_{0}^{ref}=20$ {[}deg{]} (Fig. 16
- Tab. III) are reported in Fig. 14 and Fig. 16, respectively, while
the \emph{SLL} and the $\Gamma$ values for the corresponding optimal
solutions are given in Tab. II and Tab. III, respectively. Once more
these results further support the conclusion that the \emph{PMM} gives
even more advantages in matching the reference pattern (i.e., a smaller
and smaller \emph{PM} index) (\ref{eq:_PM.Metric}) when the ratio
between the number of sub-arrays, $Q$, and the number of array elements,
$N$, reduces.

\section{\noindent Conclusions}

\noindent An innovative technique for the design of sub-arrayed phased
arrays with sub-array-only amplitude and phase controls that generate
arbitrary-shaped patterns has been presented. The synthesis problem
has been formulated in the power pattern domain and a customized synthesis
method has been developed and numerically validated. 

\noindent The main innovative contributions of this paper can be summarized
in the following ones: 

\begin{itemize}
\item the definition of an innovative paradigm for the synthesis of \emph{CPA}s
that, unlike the \emph{SoA EM} methods, directly works in the power
pattern domain where the pattern matching metric is evaluated;
\item the introduction of a customized and effective clustering method for
exploring the solution space of the \emph{EP}s and aimed at aggregating,
in the same sub-array, the array elements with similar \emph{EP}s;
\item the implementation of an innovative two-step method, the \emph{PMM},
based on a customized integration of the k-means and the \emph{IPM}
to deal with the clustering of the array and the weighting of the
sub-array clusters, respectively.
\end{itemize}
The numerical assessment has proved that: 

\begin{itemize}
\item the synthesis of \emph{CPA}s with arbitrary-shaped patterns can be
profitably and realiably addressed with the \emph{PMM}, which is generally
able to achieve the optimal clustering with the best \emph{PM} value;
\item the \emph{PMM} always outperforms the \emph{EMM} \cite{Rocca 2020}\cite{Rocca 2022}
when formulating the \emph{CPA} synthesis as a power pattern matching
problem;
\item the \emph{PMM} gets better and better than the \emph{EMM} when the
clustering ratio $\frac{Q}{N}$ reduces. 
\end{itemize}
Future works, outside the scope and objectives of this paper, will
be aimed at extending the proposed \emph{PM} approach to planar and
conformal arrays.

\section*{\noindent Appendix}

The objective of this Appendix is to prove that the summation of the
$N$ \emph{EP}s is a real quantity. 

\noindent By substituting (\ref{eq:_Elementary.Power.Pattern}) into
the right term of (\ref{eq:_FPA.Pattern.w.Elementary.Power.Patterns}),
it turns out \begin{equation}
\sum_{n=1}^{N}P_{n}\left(u\right)=\sum_{n=1}^{N}\left|AF_{n}\left(u\right)\right|^{2}+\sum_{n=1}^{N}\sum_{\ell=1,\ell\neq n}^{N}AF_{n}\left(u\right)AF_{\ell}^{*}\left(u\right).\label{eq:_Appendix.1}\end{equation}

\noindent The first summation in (\ref{eq:_Appendix.1}) is a real
value since each $n$-th ($n=1,...,N$) component, $\left|AF_{n}\left(u\right)\right|^{2}$,
is real. As for the second term, let us expand the two summations
as follows\begin{eqnarray}
\sum_{n=1}^{N}\sum_{\ell=1,\ell\neq n}^{N}AF_{n}\left(u\right)AF_{\ell}^{*}\left(u\right) & = & AF_{1}\left(u\right)AF_{2}^{*}\left(u\right)+...+AF_{n}\left(u\right)AF_{\ell}^{*}\left(u\right)+...\label{eq:_Appendix.2}\\
 &  & +AF_{\ell}\left(u\right)AF_{n}*\left(u\right)+...+AF_{N}\left(u\right)AF_{N-1}^{*}\left(u\right)\,.\nonumber \end{eqnarray}

\noindent One can notice that (\ref{eq:_Appendix.2}) includes the
sum of couple of terms, {[}$AF_{n}\left(u\right)AF_{\ell}^{*}\left(u\right)+AF_{\ell}\left(u\right)AF_{n}^{*}\left(u\right)${]},
($n,\ell\in\left[1:N\right]$, $n\neq\ell$), each providing a real
value\begin{equation}
\begin{array}{c}
AF_{n}\left(u\right)AF_{\ell}^{*}\left(u\right)+AF_{\ell}\left(u\right)AF_{n}^{*}\left(u\right)=\\
\left(\Re\left\{ AF_{n}\left(u\right)\right\} +j\Im\left\{ AF_{n}\left(u\right)\right\} \right)\times\left(\Re\left\{ AF_{\ell}\left(u\right)\right\} +j\Im\left\{ AF_{\ell}\left(u\right)\right\} \right)^{*}+\\
\left(\Re\left\{ AF_{\ell}\left(u\right)\right\} +j\Im\left\{ AF_{\ell}\left(u\right)\right\} \right)\times\left(\Re\left\{ AF_{n}\left(u\right)\right\} +j\Im\left\{ AF_{n}\left(u\right)\right\} \right)^{*}=\\
=\left(\Re\left\{ AF_{n}\left(u\right)\right\} +j\Im\left\{ AF_{n}\left(u\right)\right\} \right)\times\left(\Re\left\{ AF_{\ell}\left(u\right)\right\} -j\Im\left\{ AF_{\ell}\left(u\right)\right\} \right)+\\
\left(\Re\left\{ AF_{\ell}\left(u\right)\right\} +j\Im\left\{ AF_{\ell}\left(u\right)\right\} \right)\times\left(\Re\left\{ AF_{n}\left(u\right)\right\} -j\Im\left\{ AF_{n}\left(u\right)\right\} \right)=\\
=\Re\left\{ AF_{n}\left(u\right)\right\} \Re\left\{ AF_{\ell}\left(u\right)\right\} +j\Im\left\{ AF_{n}\left(u\right)\right\} \Re\left\{ AF_{\ell}\left(u\right)\right\} -\\
j\Re\left\{ AF_{n}\left(u\right)\right\} \Im\left\{ AF_{\ell}\left(u\right)\right\} -\Im\left\{ AF_{n}\left(u\right)\right\} \Im\left\{ AF_{\ell}\left(u\right)\right\} +\\
\Re\left\{ AF_{n}\left(u\right)\right\} \Re\left\{ AF_{\ell}\left(u\right)\right\} -j\Im\left\{ AF_{n}\left(u\right)\right\} \Re\left\{ AF_{\ell}\left(u\right)\right\} +\\
j\Re\left\{ AF_{n}\left(u\right)\right\} \Im\left\{ AF_{\ell}\left(u\right)\right\} -\Im\left\{ AF_{n}\left(u\right)\right\} \Im\left\{ AF_{\ell}\left(u\right)\right\} =\\
=2\Re\left\{ AF_{n}\left(u\right)\right\} \Re\left\{ AF_{\ell}\left(u\right)\right\} -2\Im\left\{ AF_{n}\left(u\right)\right\} \Im\left\{ AF_{\ell}\left(u\right)\right\} \,.\end{array}\label{eq:_Appendix.3}\end{equation}
Accordingly, the summation of the $N$ \emph{EP}s is a real quantity
and the equality (\ref{eq:_FPA.Pattern.w.Elementary.Power.Patterns})
is further confirmed.

\section*{\noindent Acknowledgements}

\noindent This work benefited from the networking activities carried
out within the Project AURORA - Smart Materials for Ubiquitous Energy
Harvesting, Storage, and Delivery in Next Generation Sustainable Environments
funded by the Italian Ministry for Universities and Research within
the PRIN-PNRR 2022 Program. Moreover, it benefited from the networking
activities carried out within the Project SEME@TN - Smart ElectroMagnetic
Environment in TrentiNo funded by the Autonomous Province of Trento
(CUP: C63C22000720003), the Project SPEED (Grant No. 61721001) funded
by National Science Foundation of China under the Chang-Jiang Visiting
Professorship Program, the Project Electromechanical Coupling Theory
and Design Method for Uncertain Factors of Electronic Equipment (Grant
No. 2022-JC-33) funded by Department of Science and Technology of
Shaanxi Province under the Natural Science Basic Research Program,
the Project Research on Design Method of Efficient Microwave Wireless
Energy Transfer Antenna Array for Space Power Stations' (Grant No.
2023-GHZD-35) funded by the Department of Science and Technology of
Shaanxi Province under the Key Research and Development Program, and
the Project National Centre for HPC, Big Data and Quantum Computing
(CN HPC) funded by the European Union - NextGenerationEU within the
PNRR Program (CUP: E63C22000970007). Views and opinions expressed
are however those of the author(s) only and do not necessarily reflect
those of the European Union or the European Research Council. Neither
the European Union nor the granting authority can be held responsible
for them. A. Massa wishes to thank E. Vico for her never-ending inspiration,
support, guidance, and help.

\newpage
\section*{FIGURE CAPTION}

\begin{itemize}
\item \textbf{Figure 1.} Sketch of a linear \emph{CPA}.
\item \textbf{Figure 2.} Graphical representation of the Excitation Domain,
$\mathcal{S}_{I}$, the Array Factor Domain, $\mathcal{S}_{AF}$,
and the Power Pattern Domain, $\mathcal{S}_{P}$.
\item \textbf{Figure 3.} Flowchart of the \emph{PMM}.
\item \textbf{Figure 4.} \emph{Illustrative Example} ($N=12$, $Q=8$, $d=\frac{\lambda}{2}$;
\emph{DC} pattern: $SLL^{ref}=-20$ {[}dB{]}, $\theta_{0}^{ref}=10$
{[}deg{]}) - \emph{EP} values, $\left\{ P_{n}\left(u_{m}\right);\  n=1,...,N\right\} $,
when $u_{m}=\left\{ 0.00,\ 0.25,\ 0.5,\ 0.75\right\} $.
\item \textbf{Figure 5.} \emph{Illustrative Example} ($N=12$, $Q=8$, $d=\frac{\lambda}{2}$;
\emph{DC} pattern: $SLL^{ref}=-20$ {[}dB{]}, $\theta_{0}^{ref}=10$
{[}deg{]}) - Plot of (\emph{a})-(\emph{i}) the \emph{PMM} clustering
in the power pattern domain, (\emph{e})-(\emph{h}) the corresponding
\emph{PMM} clustered array layout, and (\emph{i})-(\emph{n}) the \emph{IPM}-computed
sub-array excitations \emph{}when (\emph{a})(\emph{e})(\emph{i}) $u_{m}=0.00$,
(\emph{b})(\emph{f})(\emph{l}) $u_{m}=0.25$, (\emph{c})(\emph{g})(\emph{m})
$u_{m}=0.50$ and (\emph{d})(\emph{h})(\emph{n}) $u_{m}=0.75$.
\item \textbf{Figure 6.} \emph{Illustrative Example} ($N=12$, $Q=8$, $d=\frac{\lambda}{2}$;
\emph{DC} pattern: $SLL^{ref}=-20$ {[}dB{]}, $\theta_{0}^{ref}=10$
{[}deg{]}; $M=17$) - Plot of (\emph{a}) the \emph{PMM} power pattern
and (\emph{b}) the \emph{PM} metric, $\Gamma\left(\ \mathbf{c}_{m}^{opt},\ \mathbf{I}_{m}^{opt}\right)$,
versus the angular sample $u_{m}$ ($m=1,...,M$).
\item \textbf{Figure 7.} \emph{Illustrative Example} ($N=12$, $Q=8$, $d=\frac{\lambda}{2}$;
\emph{DC} pattern: $SLL^{ref}=-20$ {[}dB{]}, $\theta_{0}^{ref}=10$
{[}deg{]}) - Plot of the power patterns\emph{. }
\item \textbf{Figure 8.} \emph{Robustness Analysis} ($N=12$, $d=\frac{\lambda}{2}$;
$\sigma=50$) - Plot of the value of the \emph{PM} metric versus the
random seed number, $\nu$ ($\nu=1,...,\sigma$), when matching (\emph{a})(\emph{b})
a \emph{DC} pattern ($SLL^{ref}=-20$ {[}dB{]}; $\theta_{0}^{ref}=10$
{[}deg{]}) or (\emph{c})(\emph{d}) a Taylor pattern ($SLL^{ref}=-20$
{[}dB{]}, $\overline{n}=3$; $\theta_{0}^{ref}=10$ {[}deg{]}) and
setting the number of sub-arrays to (\emph{a})(\emph{c}) $Q=8$ or
(\emph{b})(\emph{d}) $Q=6$.
\item \textbf{Figure 9.} \emph{Robustness Analysis} ($N=12$, $d=\frac{\lambda}{2}$;
$M=1001$; $\sigma=50$) - Plot of the \emph{PM} metric versus the
angular samples when matching (\emph{a})(\emph{b}) a \emph{DC} pattern
($SLL^{ref}=-20$ {[}dB{]}; $\theta_{0}^{ref}=10$ {[}deg{]}) or (\emph{c})(\emph{d})
a Taylor pattern ($SLL^{ref}=-20$ {[}dB{]}, $\overline{n}=3$; $\theta_{0}^{ref}=10$
{[}deg{]}) and setting the number of sub-arrays to (\emph{a})(\emph{c})
$Q=8$ or (\emph{b})(\emph{d}) $Q=6$.
\item \textbf{Figure 10.} \emph{Comparative Assessment} ($d=\frac{\lambda}{2}$;
\emph{DC} pattern: $SLL^{ref}=-20$ {[}dB{]}, $\theta_{0}^{ref}=10$
{[}deg{]}) - Plots of the \emph{PM} metric, $\Gamma$, and the matching
improvement index, $R$, versus the number of array elements, $N$,
for a \emph{CPA} with (\emph{a}) $Q=\frac{N}{2}$ and (\emph{b}) $Q=\frac{3N}{4}$
sub-arrays.
\item \textbf{Figure 11.} \emph{Comparative Assessment} ($N=32$, $d=\frac{\lambda}{2}$;
\emph{DC} pattern: $SLL^{ref}=-20$ {[}dB{]}, $\theta_{0}^{ref}=10$
{[}deg{]}) - Plot of the power patterns for a \emph{CPA} with (\emph{a})
$Q=16$ and (\emph{b}) $Q=24$ sub-arrays.
\item \textbf{Figure 12.} \emph{Comparative Assessment} ($N=32$, $Q=16$,
$d=\frac{\lambda}{2}$; \emph{CS} pattern: $SLL^{ref}=-20$ {[}dB{]},
$RPE=1.0$ {[}dB{]}, $FNBW=40$ {[}deg{]}, $\theta_{0}^{ref}=0$ {[}deg{]})
- Plot of (\emph{a}) the power patterns, (\emph{b})(\emph{c}) the
clustering configurations in the power pattern domain, and (\emph{d})(\emph{e})
the corresponding \emph{CPA} layouts synthesized with (\emph{b})(\emph{d})
the \emph{PMM} and (\emph{c})(\emph{e}) the \emph{EMM}.
\item \textbf{Figure 13.} \emph{Comparative Assessment} ($N=32$, $d=\frac{\lambda}{2}$;
\emph{CS} pattern: $RPE=1.0$ {[}dB{]}, $FNBW=40$ {[}deg{]}, $\theta_{0}^{ref}=0$
{[}deg{]}) - Plots of the \emph{PM} metric, $\Gamma$, and of the
matching improvement index, $R$, versus the side-lobe level, $SLL^{ref}$,
for a \emph{CPA} with (\emph{a}) $Q=\frac{N}{4}$, (\emph{b}) $Q=\frac{N}{2}$,
and (\emph{c}) $Q=\frac{3N}{4}$ sub-arrays.
\item \textbf{Figure 14.} \emph{Comparative Assessment} ($N=32$, $d=\frac{\lambda}{2}$;
\emph{CS} pattern: $SLL^{ref}=-25$ {[}dB{]}, $RPE=1.0$ {[}dB{]},
$FNBW=40$ {[}deg{]}, $\theta_{0}^{ref}=0$ {[}deg{]}) - Plots of
the power patterns for a \emph{CPA} with (\emph{a}) $Q=\frac{N}{4}$,
(\emph{b}) $Q=\frac{N}{2}$, and (\emph{c}) $Q=\frac{3N}{4}$ sub-arrays.
\item \textbf{Figure 15.} \emph{Comparative Assessment} ($N=32$, $d=\frac{\lambda}{2}$;
\emph{CS} pattern: $SLL^{ref}=-20$ {[}dB{]}, $RPE=1.0$ {[}dB{]},
$FNBW=40$ {[}deg{]}) - Plots of the \emph{PM} metric, $\Gamma$,
and of the matching improvement index, $R$, versus the steering angle,
$\theta_{0}^{ref}$, for a \emph{CPA} with (\emph{a}) $Q=\frac{N}{4}$,
(\emph{b}) $Q=\frac{N}{2}$, and (\emph{c}) $Q=\frac{3N}{4}$ sub-arrays.
\item \textbf{Figure 16.} \emph{Comparative Assessment} ($N=32$, $d=\frac{\lambda}{2}$;
\emph{CS} pattern: $SLL^{ref}=-20$ {[}dB{]}, $RPE=1.0$ {[}dB{]},
$FNBW=40$ {[}deg{]}, $\theta_{0}^{ref}=20$ {[}deg{]}) - Plots of
the power patterns for a \emph{CPA} with (\emph{a}) $Q=\frac{N}{4}$,
(\emph{b}) $Q=\frac{N}{2}$, and (\emph{c}) $Q=\frac{3N}{4}$ sub-arrays.
\end{itemize}

\section*{TABLE CAPTIONS}

\begin{itemize}
\item \textbf{Table I.} \emph{Comparative Assessment} ($N=32$, $Q=16$,
$d=\frac{\lambda}{2}$; \emph{CS} pattern, $SLL^{ref}=-20$ {[}dB{]},
$RPE=1.0$ {[}dB{]}, $FNBW=40$ {[}deg{]}, $\theta_{0}^{ref}=0$ {[}deg{]})
- Pattern indexes.
\item \textbf{Table II.} \emph{Comparative Assessment} ($N=32$, $Q=\left\{ \frac{1}{4},\,\frac{1}{2},\,\frac{3}{4}\right\} \times N$,
$d=\frac{\lambda}{2}$; \emph{CS} pattern, $SLL^{ref}=-25$ {[}dB{]},
$RPE=1.0$ {[}dB{]}, $FNBW=40$ {[}deg{]}, $\theta_{0}^{ref}=0$ {[}deg{]})
- Pattern indexes.
\item \textbf{Table III.} \emph{Comparative Assessment} ($N=32$, $Q=\left\{ \frac{1}{4},\,\frac{1}{2},\,\frac{3}{4}\right\} \times N$,
$d=\frac{\lambda}{2}$; \emph{CS} pattern, $SLL^{ref}=-20$ {[}dB{]},
$RPE=1.0$ {[}dB{]}, $FNBW=40$ {[}deg{]}, $\theta_{0}^{ref}=20$
{[}deg{]}) - Pattern indexes.
\end{itemize}
\newpage
~\vfill

\begin{center}\begin{tabular}{c}
\includegraphics[%
  width=1.0\columnwidth]{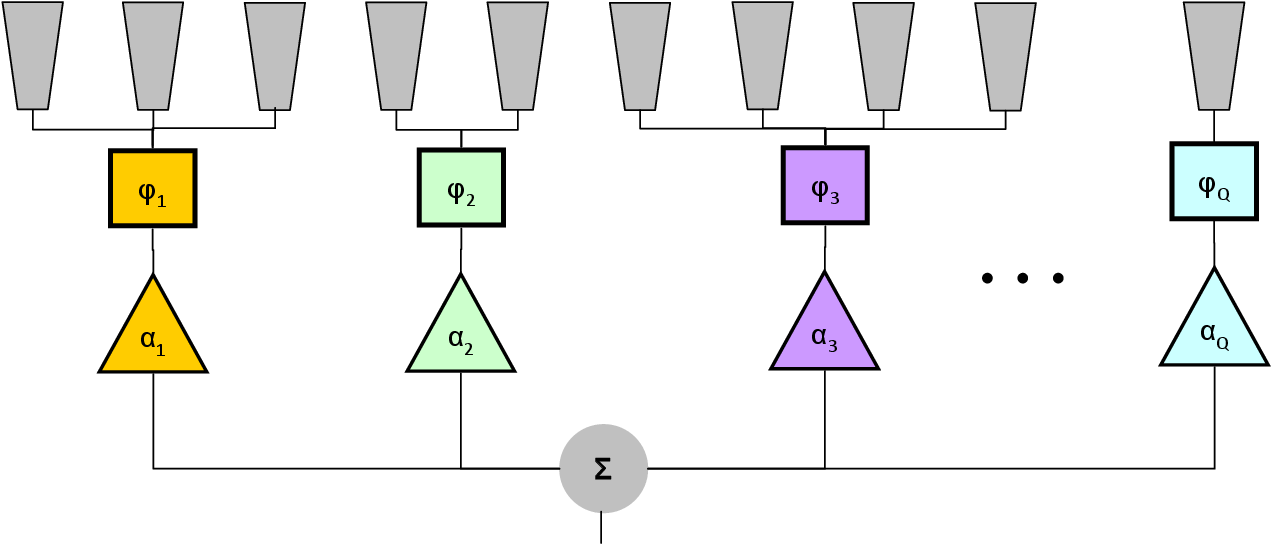}\tabularnewline
\end{tabular}\end{center}

\begin{center}~\vfill\end{center}

\begin{center}\textbf{Fig. 1 - A. Benoni} \textbf{\emph{et al.}}\textbf{,}
\textbf{\emph{{}``}}Design of Clustered ...''\end{center}

\newpage
~\vfill

\begin{center}\begin{tabular}{c}
\includegraphics[%
  width=1.0\columnwidth]{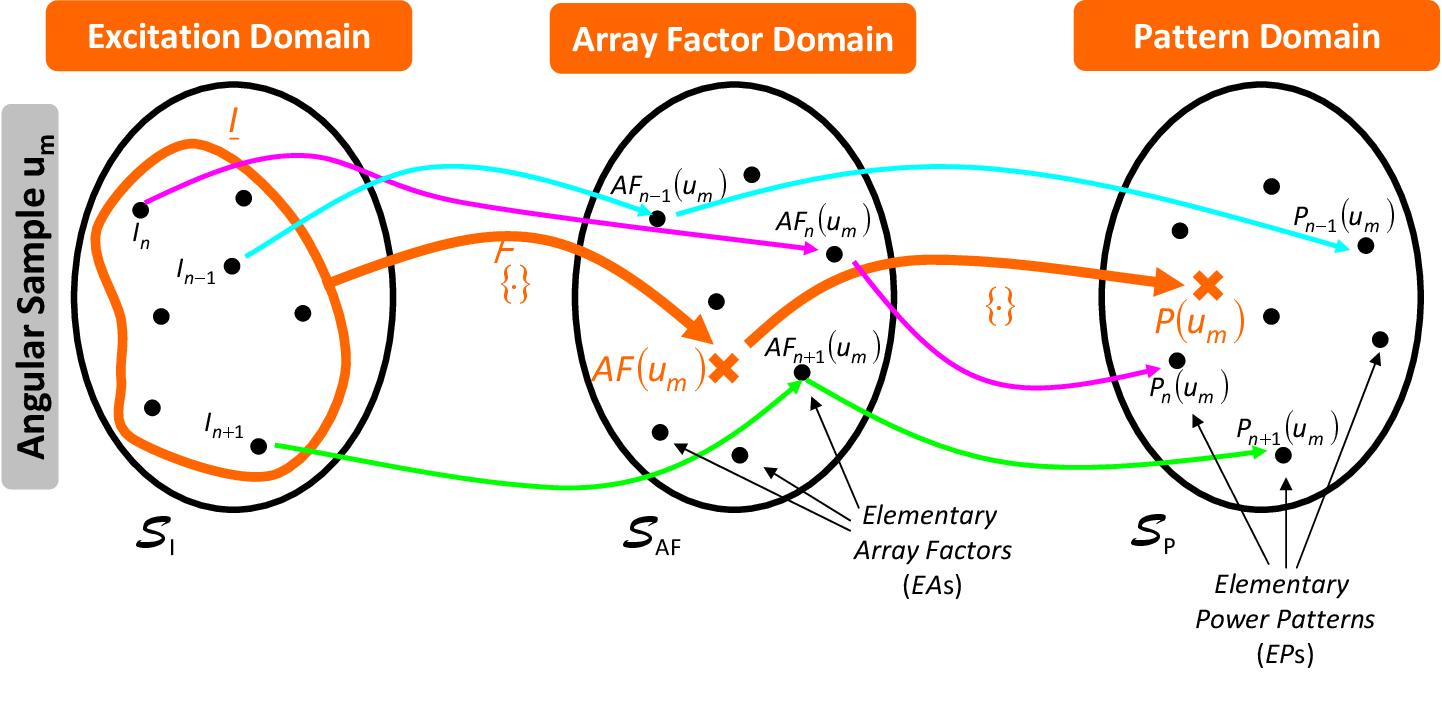}\tabularnewline
\end{tabular}\end{center}

\begin{center}~\vfill\end{center}

\begin{center}\textbf{Fig. 2 - A. Benoni} \textbf{\emph{et al.}}\textbf{,}
\textbf{\emph{{}``}}Design of Clustered ...''\end{center}

\newpage
~\vfill

\begin{center}\begin{tabular}{c}
\includegraphics[%
  width=1.0\columnwidth]{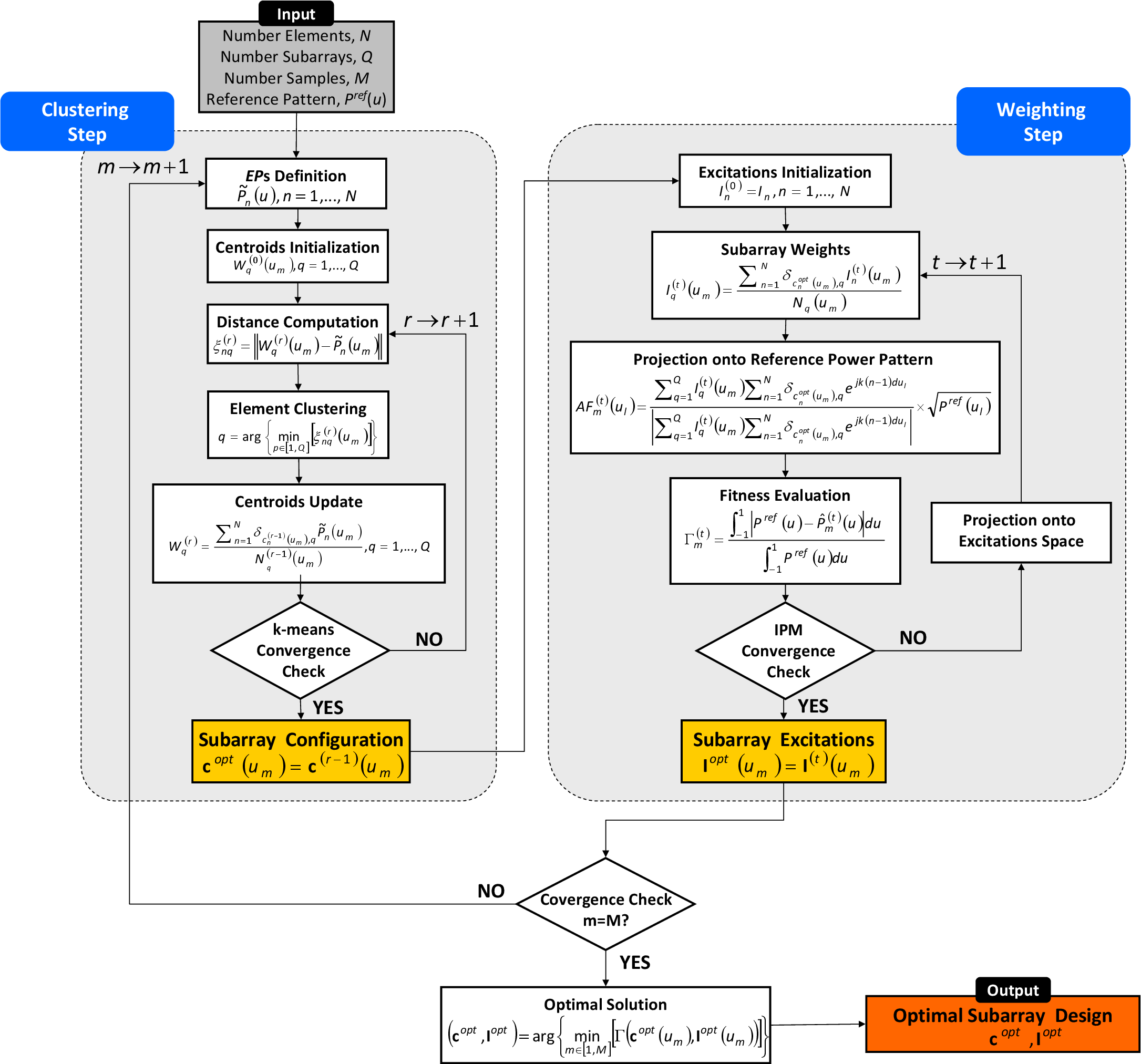}\tabularnewline
\end{tabular}\end{center}

\begin{center}~\vfill\end{center}

\begin{center}\textbf{Fig. 3 - A. Benoni} \textbf{\emph{et al.}}\textbf{,}
\textbf{\emph{{}``}}Design of Clustered ...''\end{center}

\newpage
~\vfill

\begin{center}\begin{tabular}{c}
\includegraphics[%
  width=0.70\columnwidth]{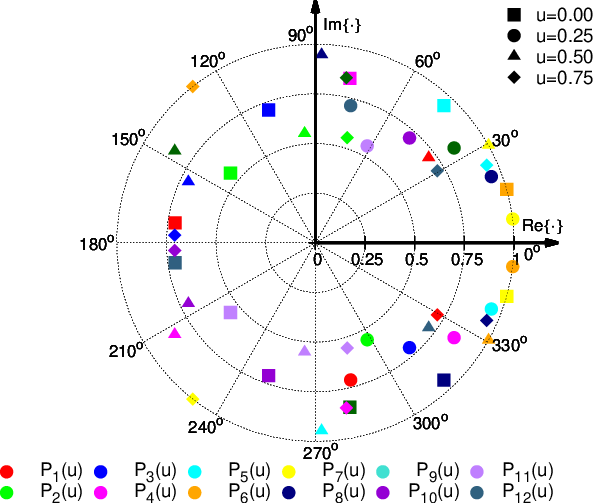}\tabularnewline
\end{tabular}\end{center}

\begin{center}~\vfill\end{center}

\begin{center}\textbf{Fig. 4 - A. Benoni} \textbf{\emph{et al.}}\textbf{,}
\textbf{\emph{{}``}}Design of Clustered ...''\end{center}

\newpage
~\vfill

\begin{center}\begin{tabular}{ccc}
\includegraphics[%
  width=0.28\columnwidth]{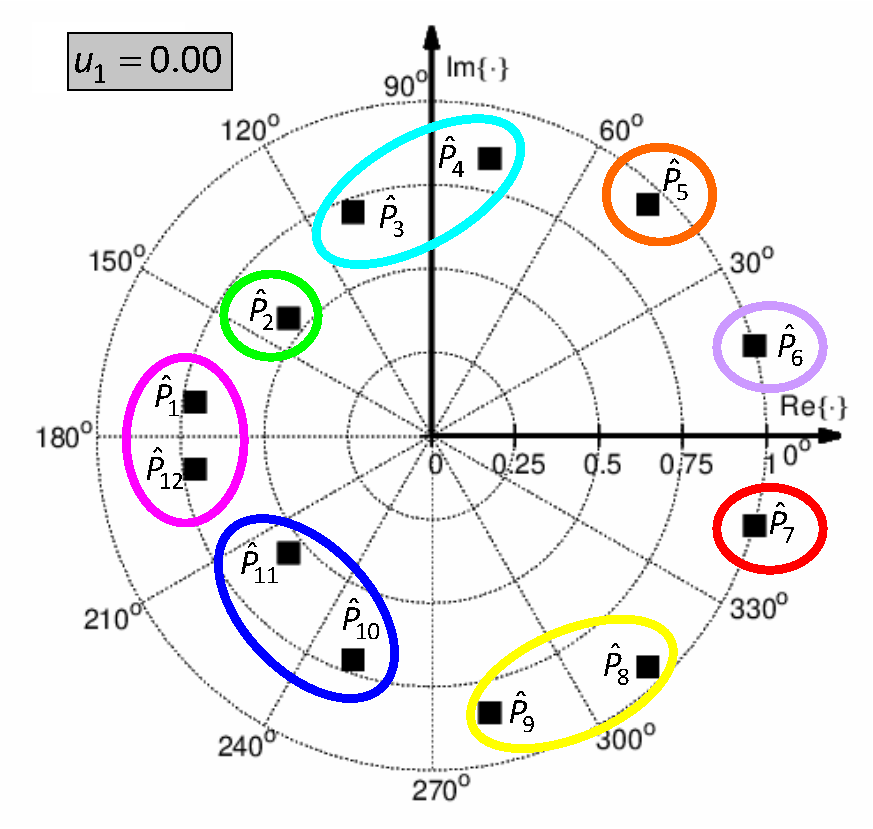}&
\includegraphics[%
  width=0.25\columnwidth]{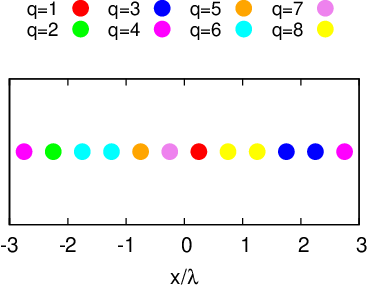}&
\includegraphics[%
  width=0.33\columnwidth]{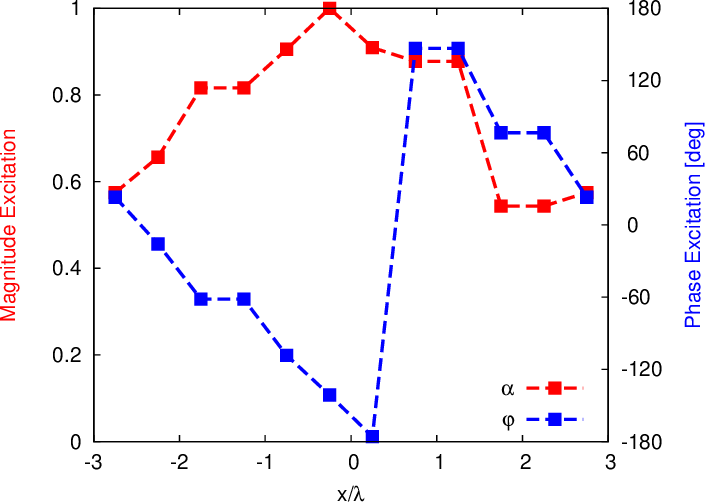}\tabularnewline
(\emph{a})&
(\emph{e})&
(\emph{i})\tabularnewline
\includegraphics[%
  width=0.28\columnwidth]{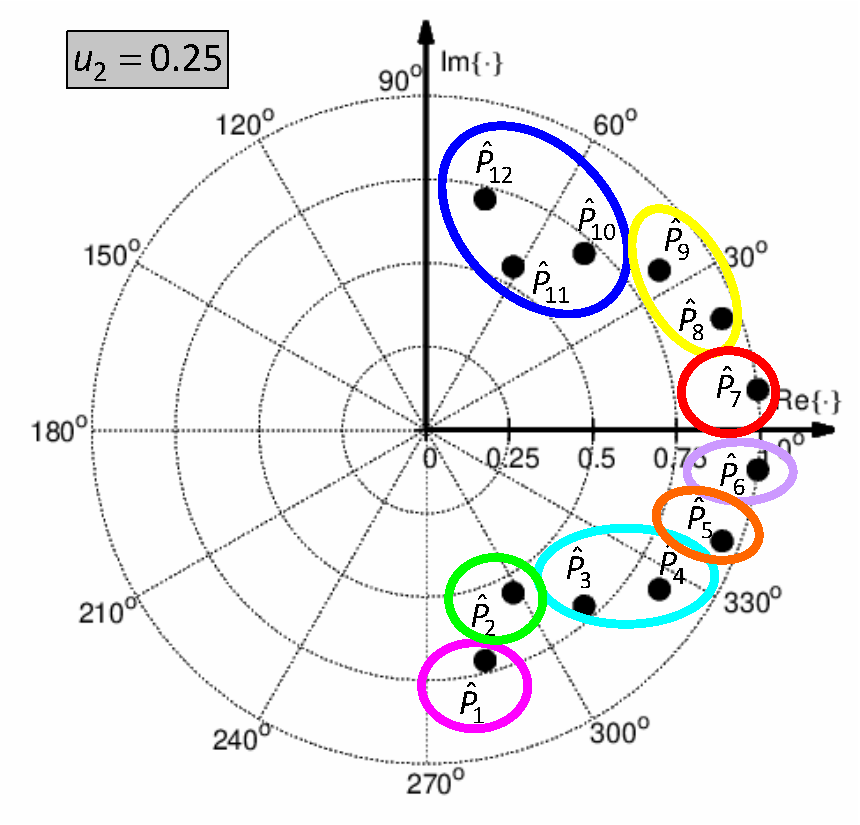}&
\includegraphics[%
  width=0.25\columnwidth]{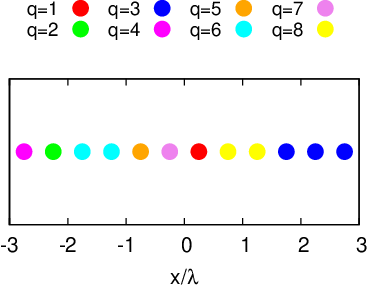}&
\includegraphics[%
  width=0.33\columnwidth]{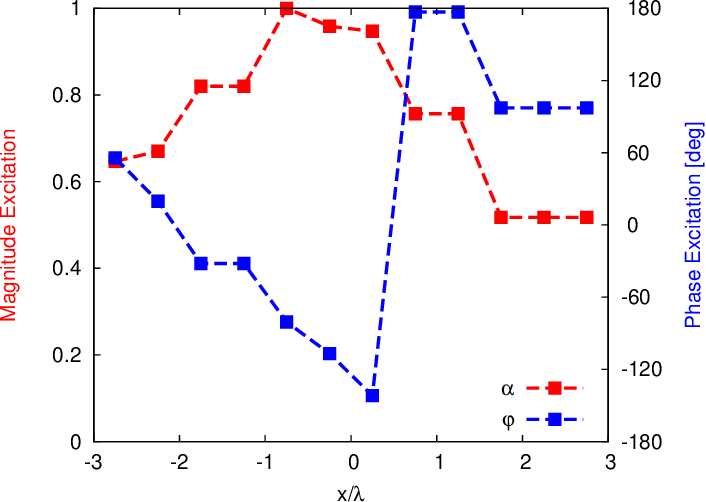}\tabularnewline
(\emph{b})&
(\emph{f})&
(\emph{l})\tabularnewline
\includegraphics[%
  width=0.28\columnwidth]{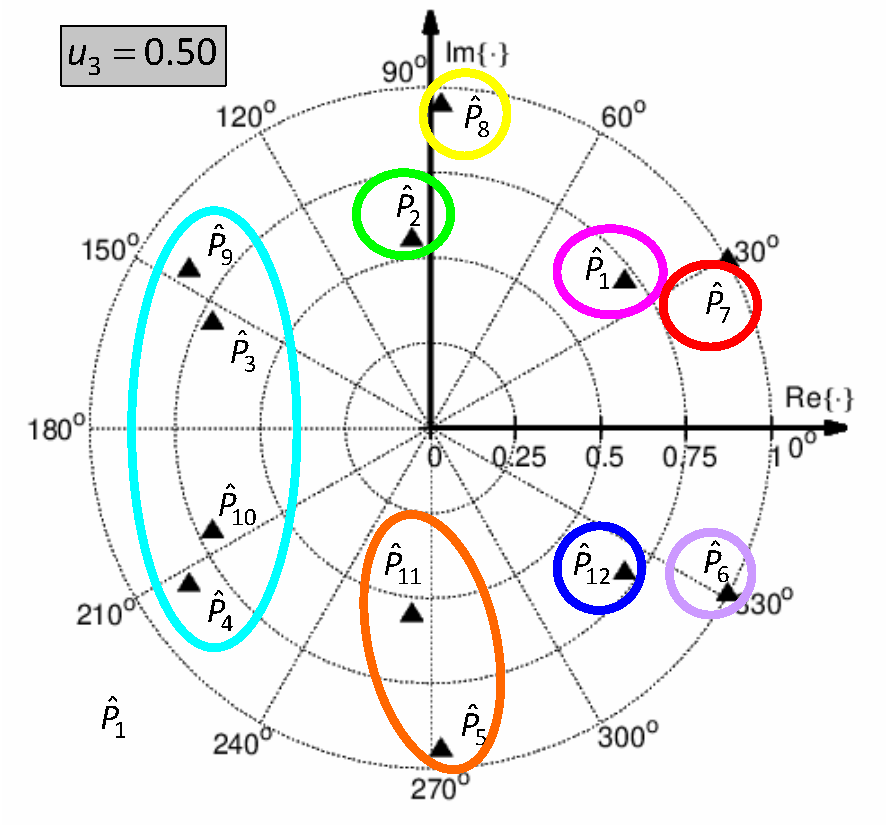}&
\includegraphics[%
  width=0.25\columnwidth]{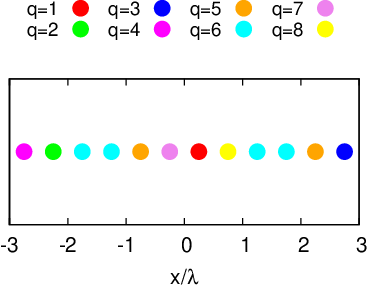}&
\includegraphics[%
  width=0.33\columnwidth]{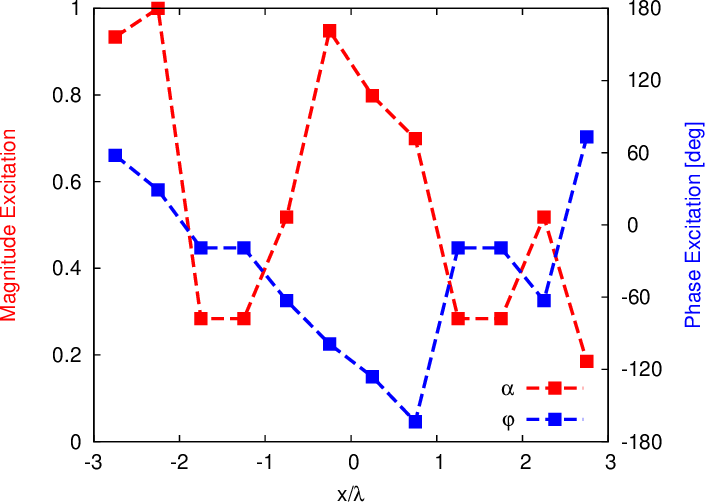}\tabularnewline
(\emph{c})&
(\emph{g})&
(\emph{m})\tabularnewline
\includegraphics[%
  width=0.28\columnwidth]{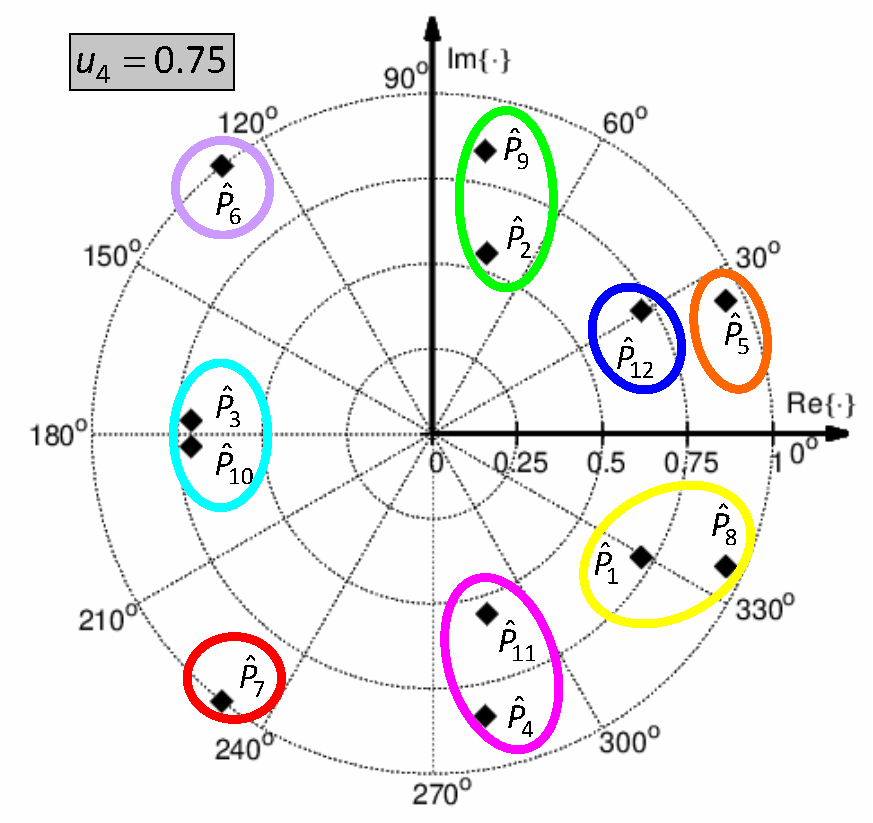}&
\includegraphics[%
  width=0.25\columnwidth]{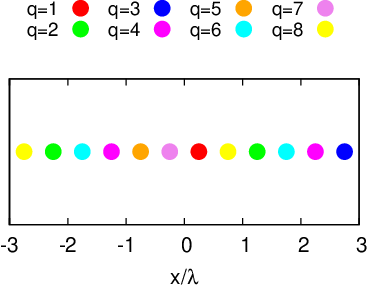}&
\includegraphics[%
  width=0.33\columnwidth]{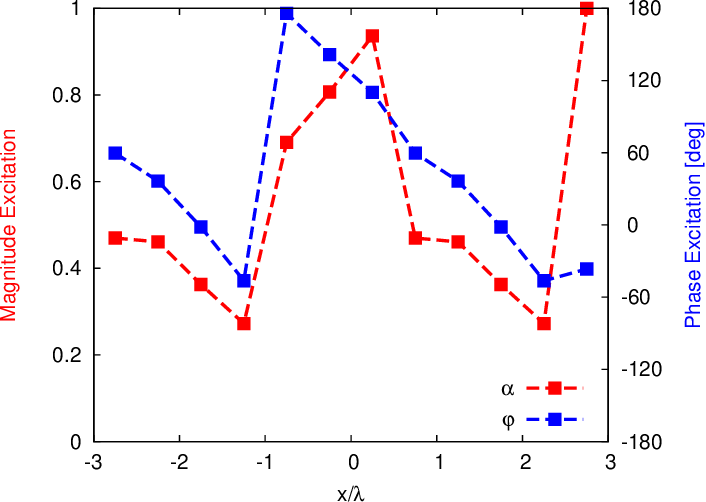}\tabularnewline
(\emph{d})&
(\emph{h})&
(\emph{n})\tabularnewline
\end{tabular}\end{center}

\begin{center}~\vfill\end{center}

\begin{center}\textbf{Fig. 5 - A. Benoni} \textbf{\emph{et al.}}\textbf{,}
\textbf{\emph{{}``}}Design of Clustered ...''\end{center}

\newpage
~\vfill

\begin{center}\begin{tabular}{c}
\includegraphics[%
  width=0.70\columnwidth]{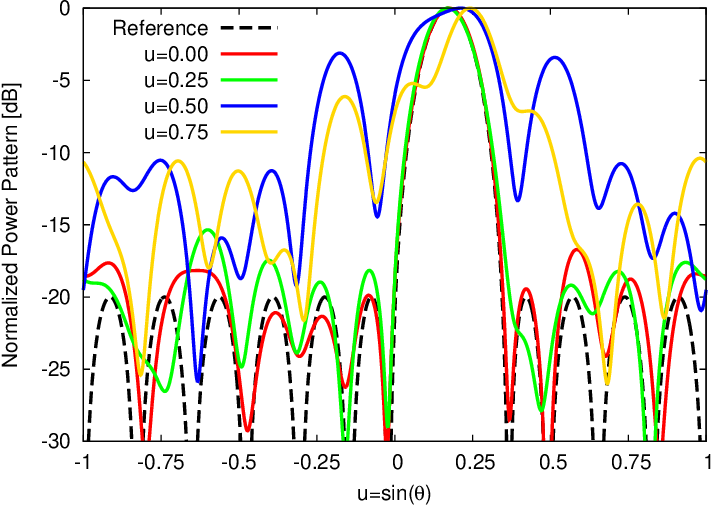}\tabularnewline
(\emph{a})\tabularnewline
\includegraphics[%
  width=0.70\columnwidth]{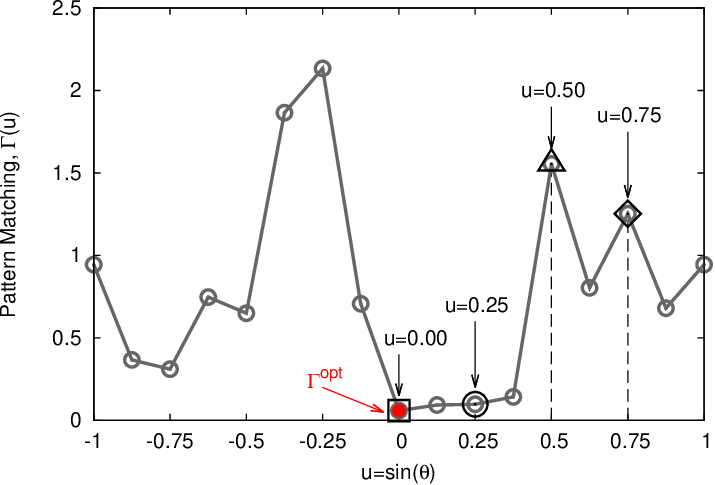}\tabularnewline
(\emph{b})\tabularnewline
\end{tabular}\end{center}

\begin{center}~\vfill\end{center}

\begin{center}\textbf{Fig. 6 - A. Benoni} \textbf{\emph{et al.}}\textbf{,}
\textbf{\emph{{}``}}Design of Clustered ...''\end{center}

\newpage
~\vfill

\begin{center}\begin{tabular}{c}
\includegraphics[%
  width=0.75\columnwidth]{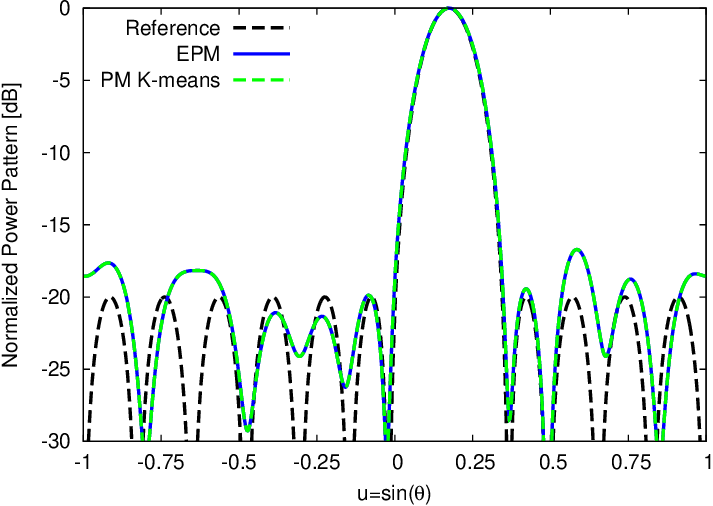}\tabularnewline
\end{tabular}\end{center}

\begin{center}~\vfill\end{center}

\begin{center}\textbf{Fig. 7 - A. Benoni} \textbf{\emph{et al.}}\textbf{,}
\textbf{\emph{{}``}}Design of Clustered ...''\end{center}

\newpage
~\vfill

\begin{center}\begin{tabular}{cc}
\includegraphics[%
  width=0.50\columnwidth]{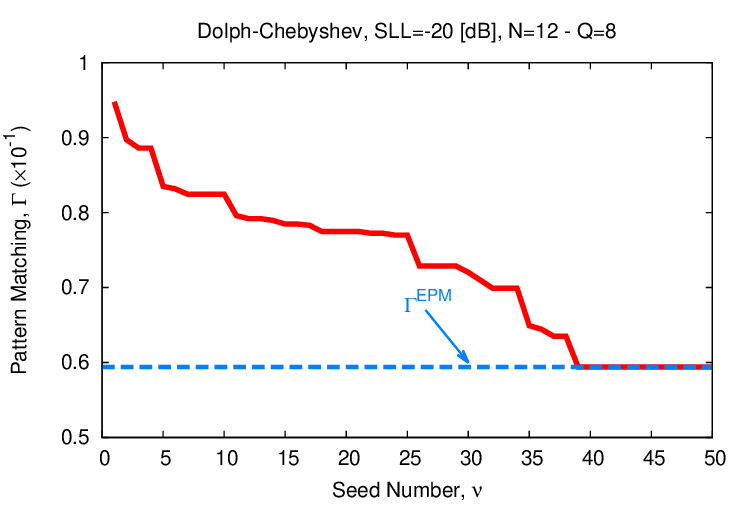}&
\includegraphics[%
  width=0.50\columnwidth]{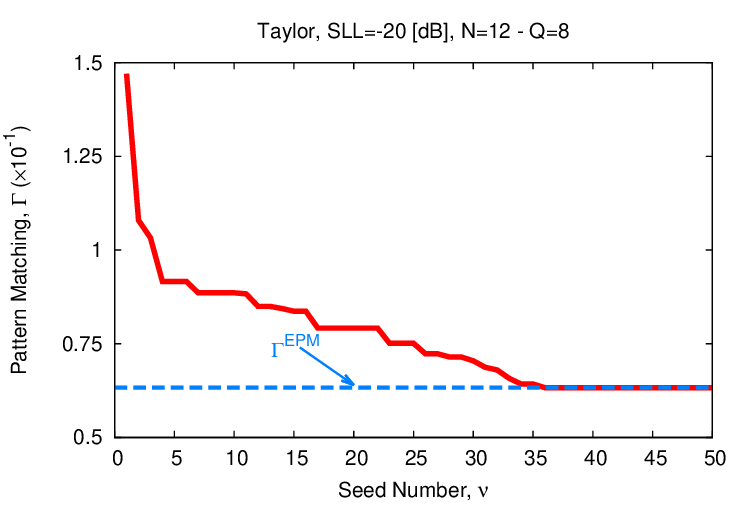}\tabularnewline
(\emph{a})&
(\emph{b})\tabularnewline
\includegraphics[%
  width=0.50\columnwidth]{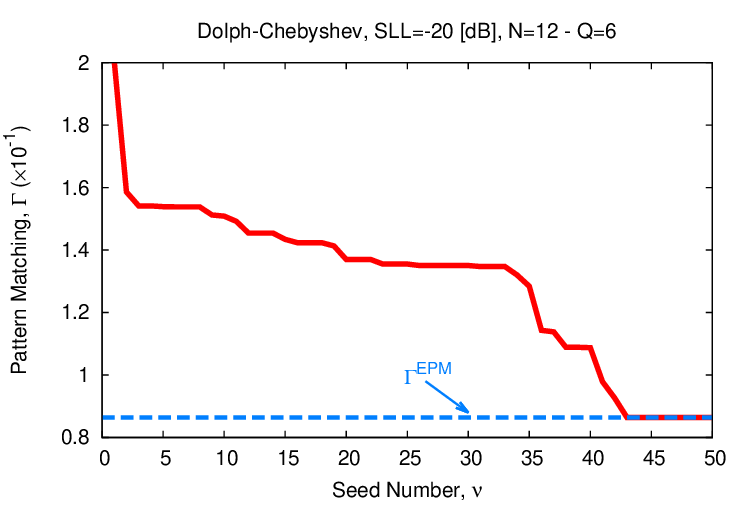}&
\includegraphics[%
  width=0.50\columnwidth]{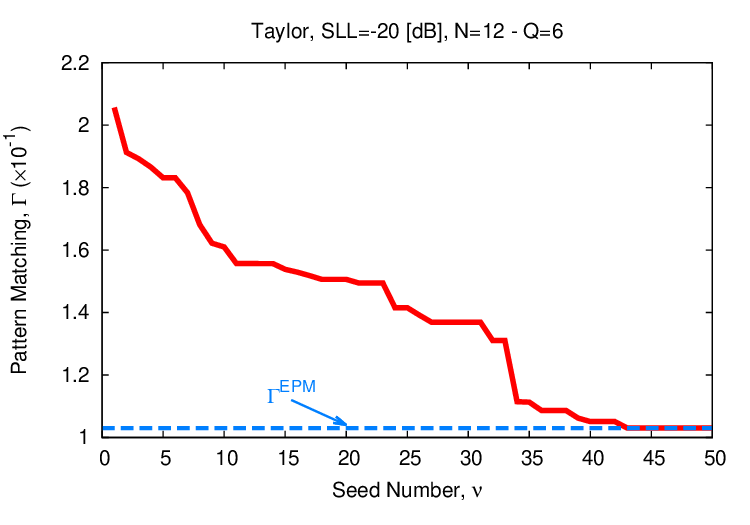}\tabularnewline
(\emph{c})&
(\emph{d})\tabularnewline
\end{tabular}\end{center}

\begin{center}~\vfill\end{center}

\begin{center}\textbf{Fig. 8 - A. Benoni} \textbf{\emph{et al.}}\textbf{,}
\textbf{\emph{{}``}}Design of Clustered ...''\end{center}

\newpage
~\vfill

\begin{center}\begin{tabular}{cc}
\includegraphics[%
  width=0.50\columnwidth]{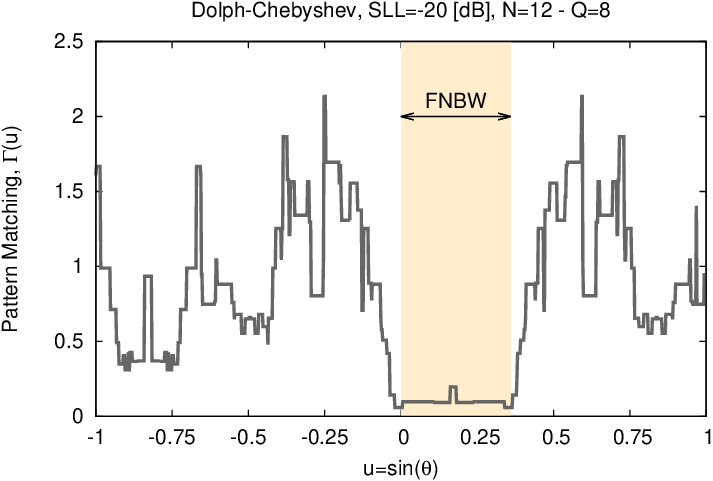}&
\includegraphics[%
  width=0.50\columnwidth]{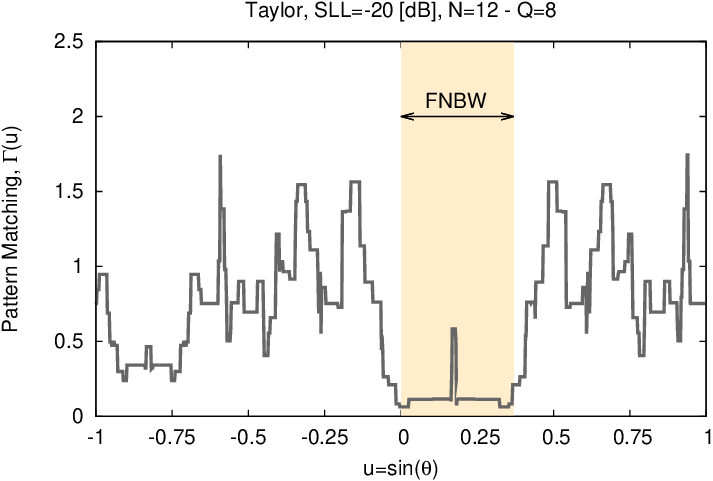}\tabularnewline
(\emph{a})&
(\emph{b})\tabularnewline
\includegraphics[%
  width=0.50\columnwidth]{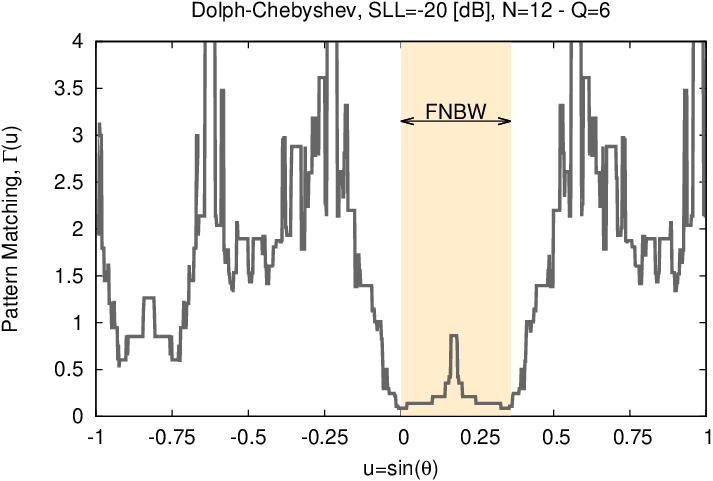}&
\includegraphics[%
  width=0.50\columnwidth]{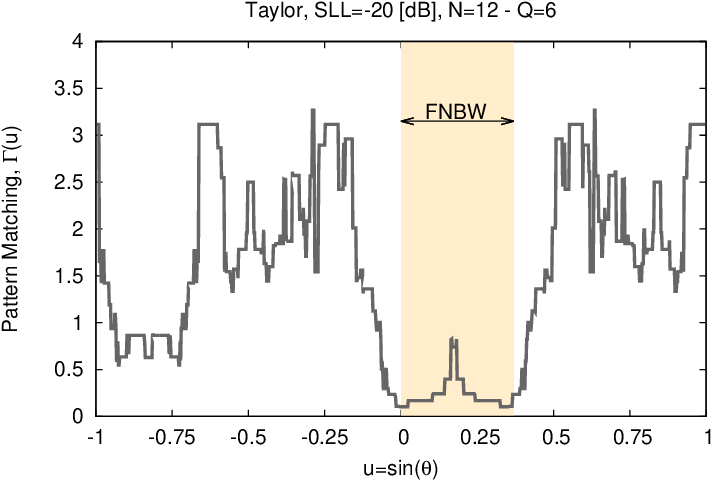}\tabularnewline
(\emph{c})&
(\emph{d})\tabularnewline
\end{tabular}\end{center}

\begin{center}~\vfill\end{center}

\begin{center}\textbf{Fig. 9 - A. Benoni} \textbf{\emph{et al.}}\textbf{,}
\textbf{\emph{{}``}}Design of Clustered ...''\end{center}

\newpage
~\vfill

\begin{center}\begin{tabular}{c}
\includegraphics[%
  width=0.70\columnwidth]{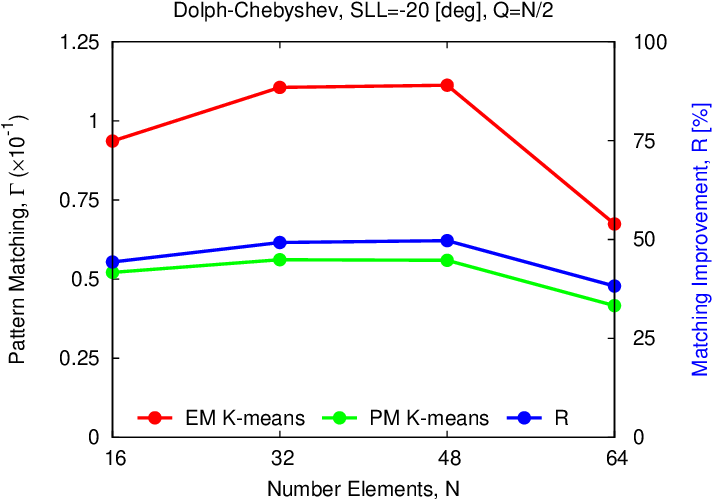}\tabularnewline
(\emph{a})\tabularnewline
\includegraphics[%
  width=0.70\columnwidth]{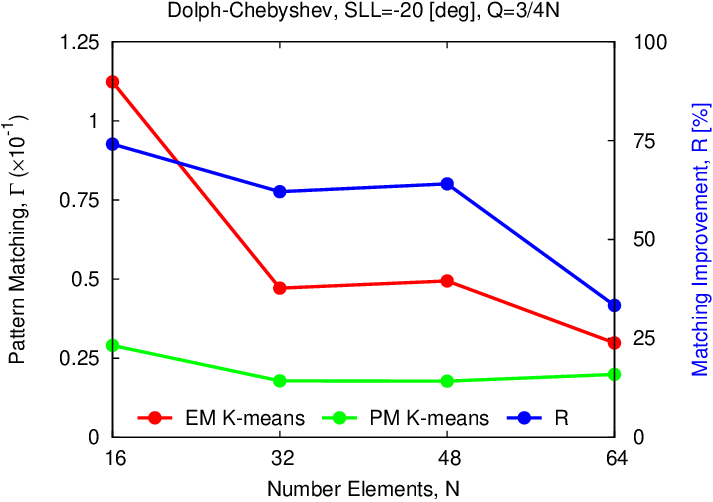}\tabularnewline
(\emph{b})\tabularnewline
\end{tabular}\end{center}

\begin{center}~\vfill\end{center}

\begin{center}\textbf{Fig. 10 - A. Benoni} \textbf{\emph{et al.}}\textbf{,}
\textbf{\emph{{}``}}Design of Clustered ...''\end{center}

\newpage
~\vfill

\begin{center}\begin{tabular}{c}
\includegraphics[%
  width=0.70\columnwidth]{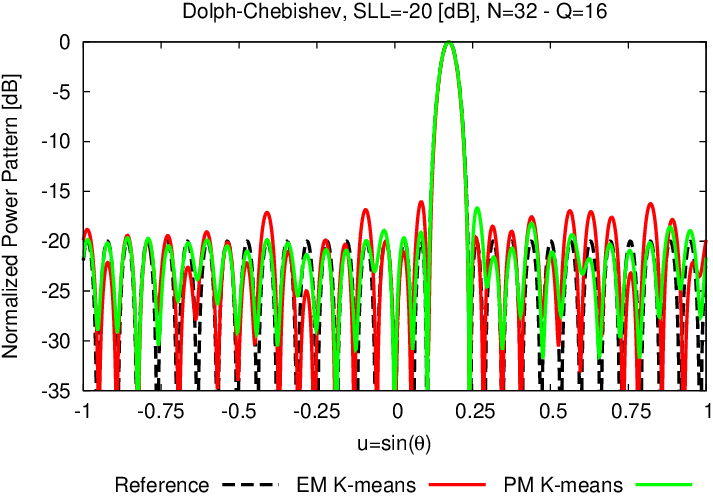}\tabularnewline
(\emph{a})\tabularnewline
\includegraphics[%
  width=0.70\columnwidth]{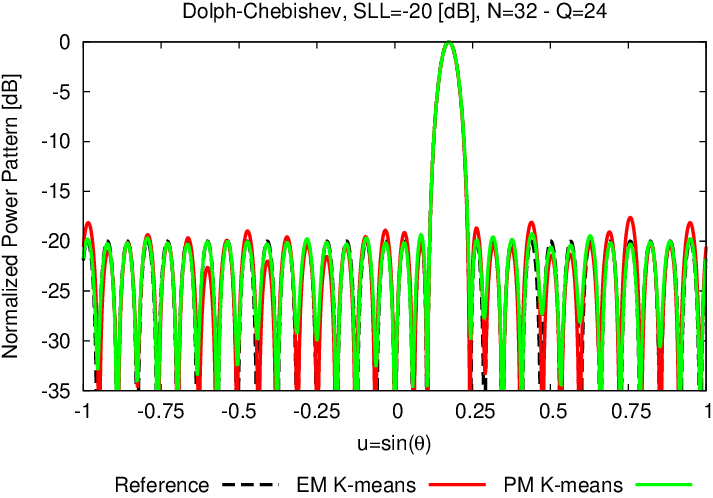}\tabularnewline
(\emph{b})\tabularnewline
\end{tabular}\end{center}

\begin{center}~\vfill\end{center}

\begin{center}\textbf{Fig. 11 - A. Benoni} \textbf{\emph{et al.}}\textbf{,}
\textbf{\emph{{}``}}Design of Clustered ...''\end{center}

\newpage
~\vfill

\begin{center}\begin{tabular}{cc}
\multicolumn{2}{c}{\includegraphics[%
  width=0.70\columnwidth]{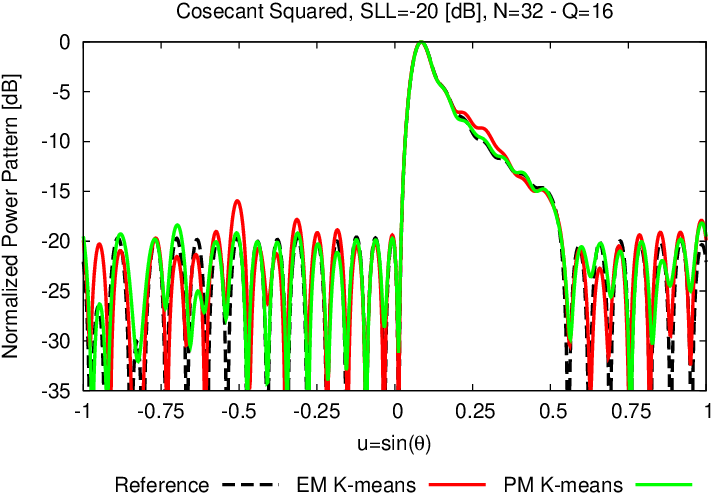}}\tabularnewline
\multicolumn{2}{c}{(\emph{a})}\tabularnewline
\includegraphics[%
  width=0.45\columnwidth]{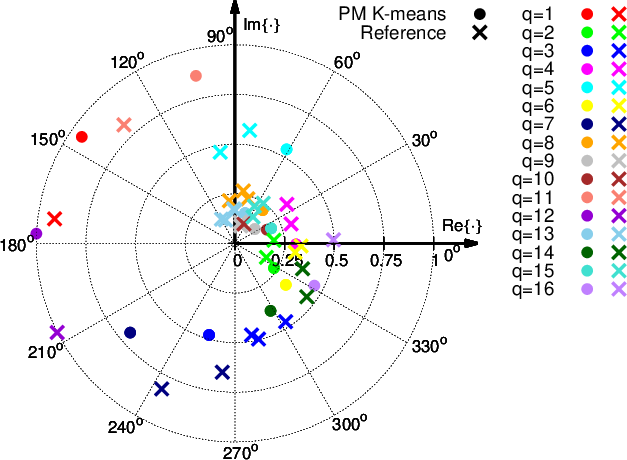}&
\includegraphics[%
  width=0.45\columnwidth]{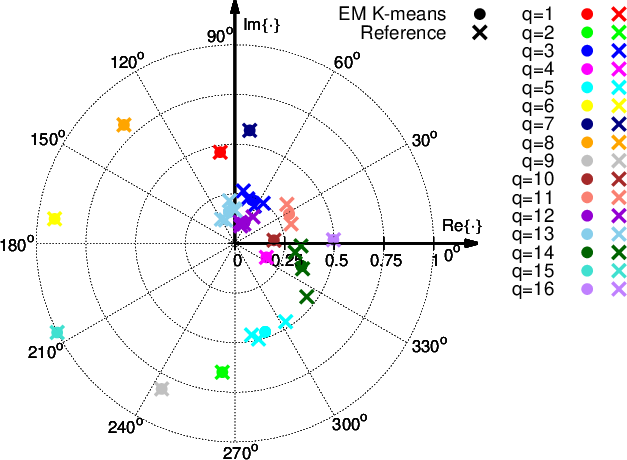}\tabularnewline
(\emph{b})&
(\emph{c})\tabularnewline
\includegraphics[%
  width=0.50\columnwidth]{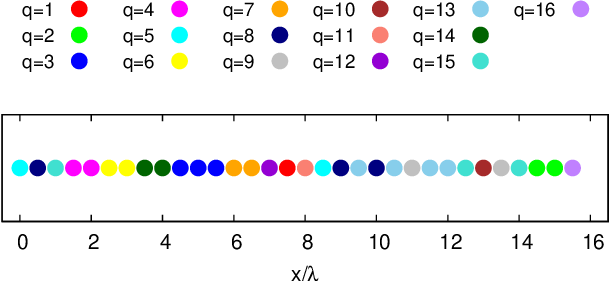}&
\includegraphics[%
  width=0.50\columnwidth]{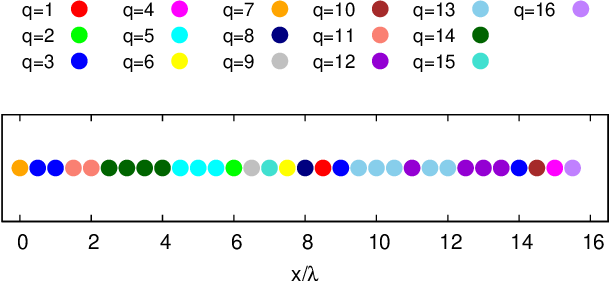}\tabularnewline
(\emph{d})&
(\emph{e})\tabularnewline
\end{tabular}\end{center}

\begin{center}~\vfill\end{center}

\begin{center}\textbf{Fig. 12 - A. Benoni} \textbf{\emph{et al.}}\textbf{,}
\textbf{\emph{{}``}}Design of Clustered ...''\end{center}

\newpage
~\vfill

\begin{center}\begin{tabular}{c}
\includegraphics[%
  width=0.50\columnwidth]{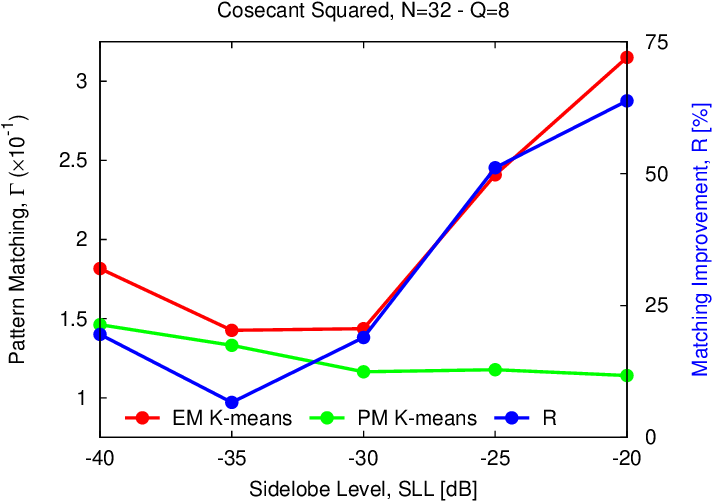}\tabularnewline
(\emph{a})\tabularnewline
\includegraphics[%
  width=0.50\columnwidth]{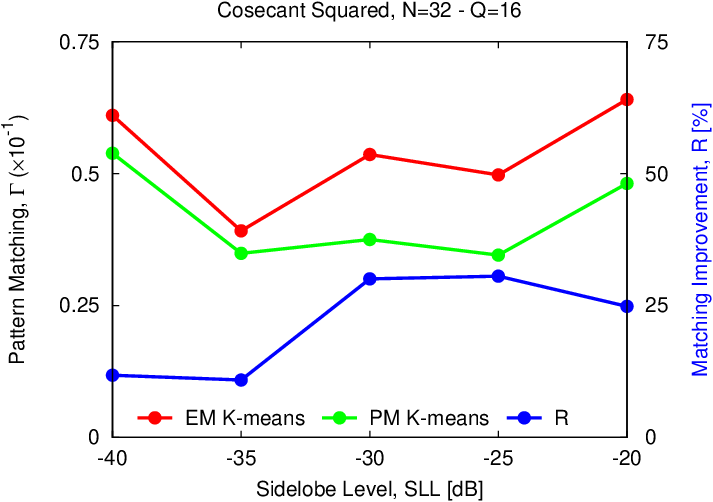}\tabularnewline
(\emph{b})\tabularnewline
\includegraphics[%
  width=0.50\columnwidth]{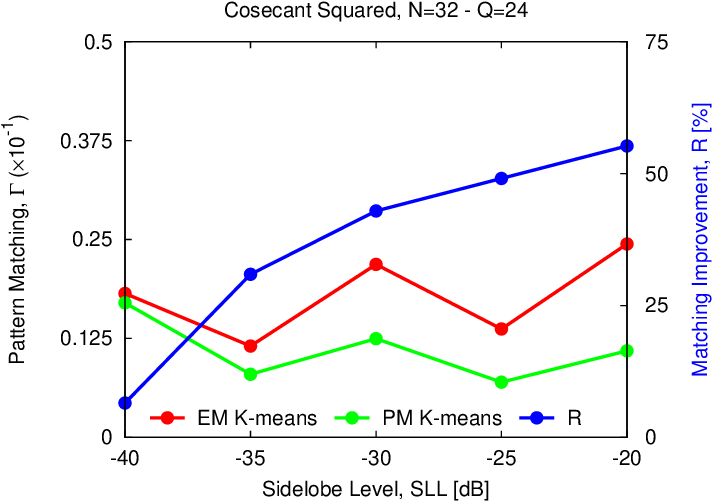}\tabularnewline
(\emph{c})\tabularnewline
\end{tabular}\end{center}

\begin{center}~\vfill\end{center}

\begin{center}\textbf{Fig. 13 - A. Benoni} \textbf{\emph{et al.}}\textbf{,}
\textbf{\emph{{}``}}Design of Clustered ...''\end{center}

\newpage
~\vfill

\begin{center}\begin{tabular}{c}
\includegraphics[%
  width=0.50\columnwidth]{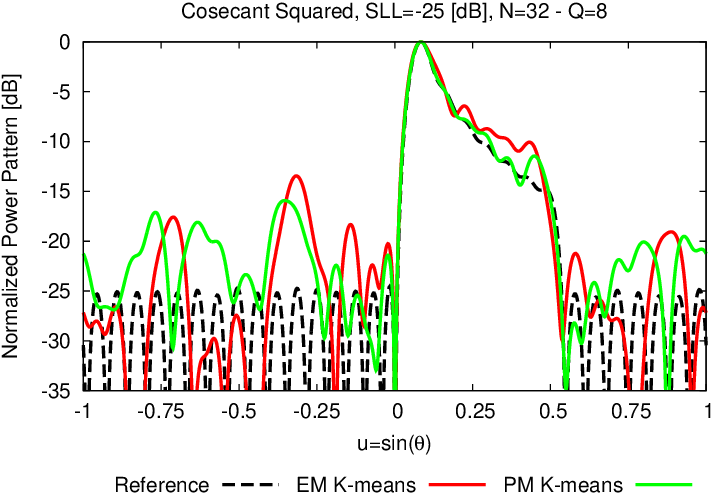}\tabularnewline
(\emph{a})\tabularnewline
\includegraphics[%
  width=0.50\columnwidth]{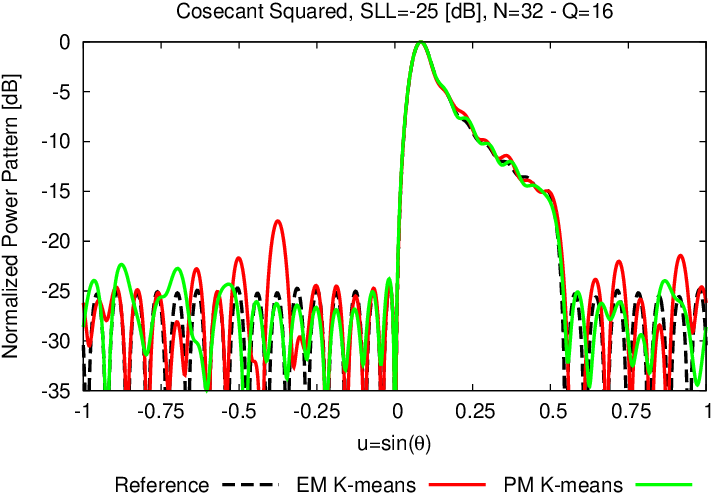}\tabularnewline
(\emph{b})\tabularnewline
\includegraphics[%
  width=0.50\columnwidth]{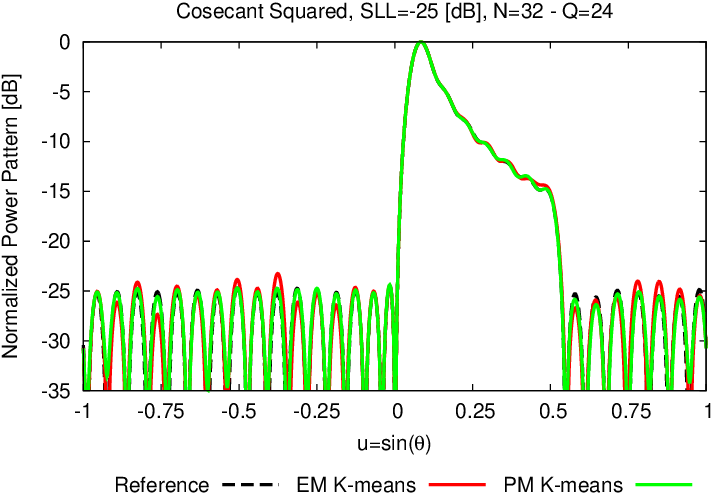}\tabularnewline
(\emph{c})\tabularnewline
\end{tabular}\end{center}

\begin{center}~\vfill\end{center}

\begin{center}\textbf{Fig. 14 - A. Benoni} \textbf{\emph{et al.}}\textbf{,}
\textbf{\emph{{}``}}Design of Clustered ...''\end{center}

\newpage
~\vfill

\begin{center}\begin{tabular}{c}
\includegraphics[%
  width=0.50\columnwidth]{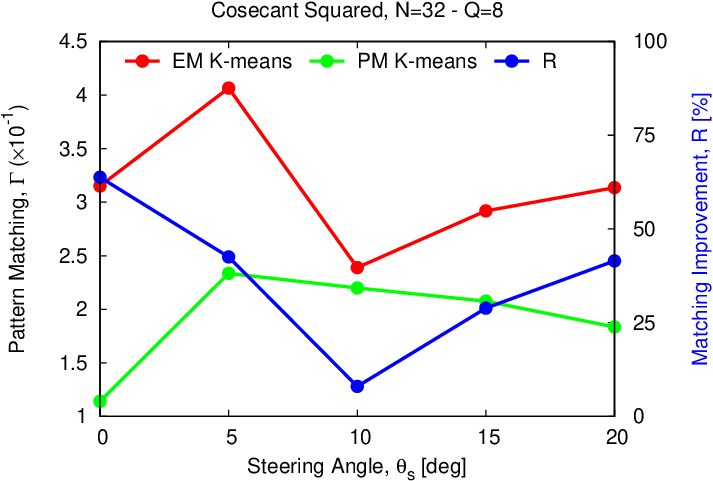}\tabularnewline
(\emph{a})\tabularnewline
\includegraphics[%
  width=0.50\columnwidth]{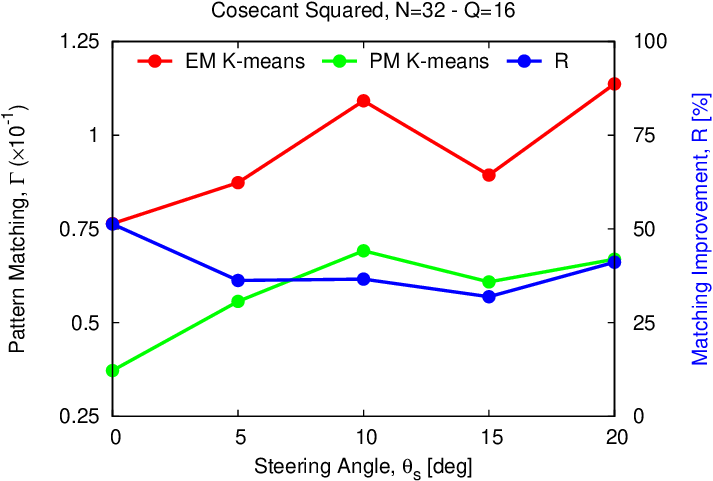}\tabularnewline
(\emph{b})\tabularnewline
\includegraphics[%
  width=0.50\columnwidth]{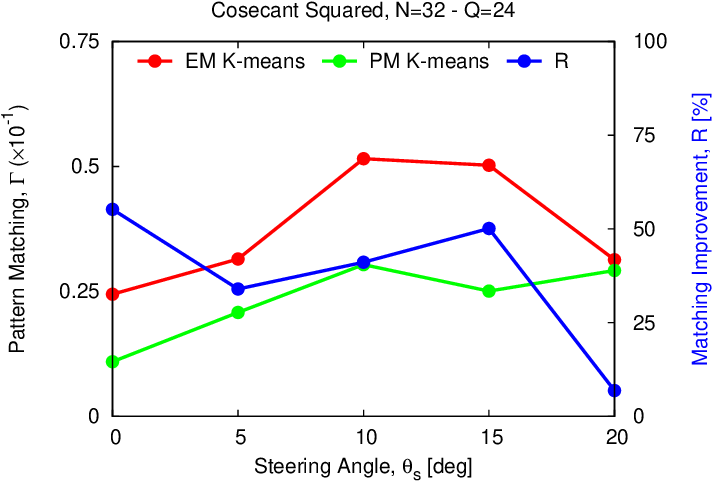}\tabularnewline
(\emph{c})\tabularnewline
\end{tabular}\end{center}

\begin{center}~\vfill\end{center}

\begin{center}\textbf{Fig. 15 - A. Benoni} \textbf{\emph{et al.}}\textbf{,}
\textbf{\emph{{}``}}Design of Clustered ...''\end{center}

\newpage
~\vfill

\begin{center}\begin{tabular}{c}
\includegraphics[%
  width=0.50\columnwidth]{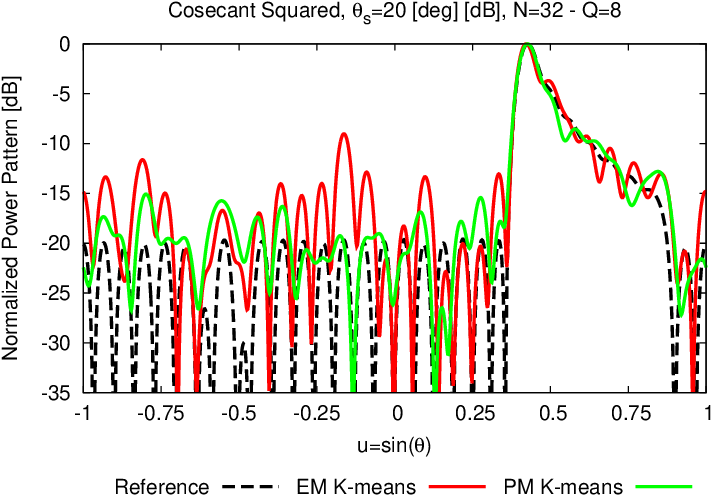}\tabularnewline
(\emph{a})\tabularnewline
\includegraphics[%
  width=0.50\columnwidth]{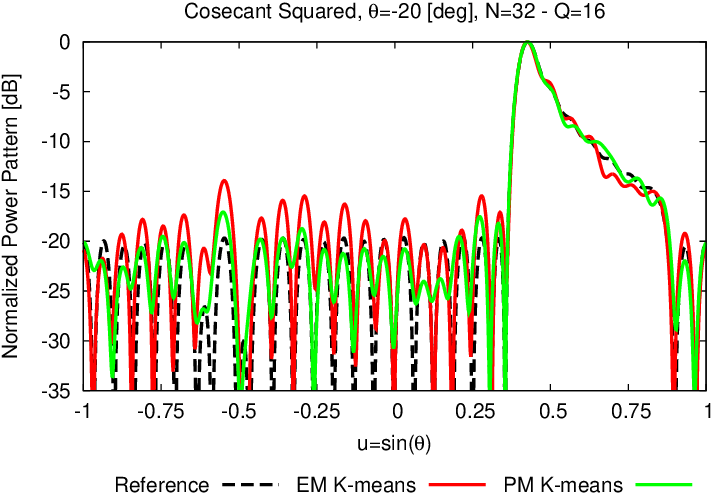}\tabularnewline
(\emph{b})\tabularnewline
\includegraphics[%
  width=0.50\columnwidth]{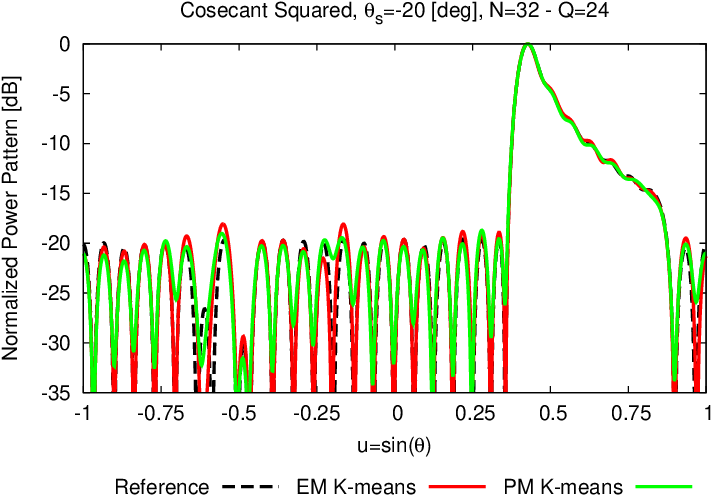}\tabularnewline
(\emph{c})\tabularnewline
\end{tabular}\end{center}

\begin{center}~\vfill\end{center}

\begin{center}\textbf{Fig. 16 - A. Benoni} \textbf{\emph{et al.}}\textbf{,}
\textbf{\emph{{}``}}Design of Clustered ...''\end{center}

\newpage
~\vfill

\begin{center}\begin{tabular}{|c||c|c|}
\hline 
\emph{Solution}&
\textbf{$SLL$} {[}dB{]}&
\textbf{$\Gamma^{opt}$}\tabularnewline
\hline
\hline 
Reference&
$-20.00$&
$-$\tabularnewline
\hline
\hline 
PMM &
$-18.22$&
$3.72\times10^{-2}$\tabularnewline
\hline 
EMM&
$-15.97$&
$7.64\times10^{-2}$\tabularnewline
\hline
\end{tabular}\end{center}

\begin{center}~\vfill\end{center}

\begin{center}\textbf{Tab. I - A. Benoni} \textbf{\emph{et al.}}\textbf{,}
\textbf{\emph{{}``}}Design of Clustered ...''\end{center}

\newpage
~\vfill

\begin{center}\begin{tabular}{|c|c||c|c|}
\hline 
\emph{Solution}&
$Q$&
\textbf{$SLL$} {[}dB{]}&
\textbf{$\Gamma^{opt}$}\tabularnewline
\hline
\hline 
Reference&
$-$&
$-25.00$&
$-$\tabularnewline
\hline
\hline 
PMM &
$8$&
$-15.93$&
$1.18\times10^{-1}$\tabularnewline
\hline 
EMM&
$8$&
$-13.52$&
$2.41\times10^{-1}$\tabularnewline
\hline
\hline 
PMM &
$16$&
$-22.33$&
$3.45\times10^{-2}$\tabularnewline
\hline 
EMM&
$16$&
$-17.97$&
$4.97\times10^{-2}$\tabularnewline
\hline
\hline 
PMM &
$24$&
$-24.39$&
$6.97\times10^{-3}$\tabularnewline
\hline 
EMM&
$24$&
$-23.26$&
$1.37\times10^{-2}$\tabularnewline
\hline
\end{tabular}\end{center}

\begin{center}~\vfill\end{center}

\begin{center}\textbf{Tab. II - A. Benoni} \textbf{\emph{et al.}}\textbf{,}
\textbf{\emph{{}``}}Design of Clustered ...''\end{center}

\newpage
~\vfill

\begin{center}\begin{tabular}{|c|c||c|c|}
\hline 
\emph{Solution}&
$Q$&
\textbf{$SLL$} {[}dB{]}&
\textbf{$\Gamma^{opt}$}\tabularnewline
\hline
\hline 
Reference&
$-$&
$-20.00$&
$-$\tabularnewline
\hline
\hline 
PMM &
$8$&
$-15.13$&
$1.83\times10^{-1}$\tabularnewline
\hline 
EMM&
$8$&
$-9.08$&
$3.13\times10^{-1}$\tabularnewline
\hline
\hline 
PMM &
$16$&
$-17.16$&
$6.69\times10^{-2}$\tabularnewline
\hline 
EMM&
$16$&
$-14.04$&
$1.13\times10^{-1}$\tabularnewline
\hline
\hline 
PMM &
$24$&
$-18.74$&
$2.92\times10^{-2}$\tabularnewline
\hline 
EMM&
$24$&
$-18.12$&
$3.14\times10^{-2}$\tabularnewline
\hline
\end{tabular}\end{center}

\begin{center}~\vfill\end{center}

\begin{center}\textbf{Tab. III - A. Benoni} \textbf{\emph{et al.}}\textbf{,}
\textbf{\emph{{}``}}Design of Clustered ...''\end{center}
\end{document}